\documentclass[11pt,a4paper]{article}
\pdfoutput=1
\usepackage{jheppub}
\usepackage[english]{babel}
\usepackage{amsfonts,amsmath,amssymb,amsthm,mathtools}
\usepackage{comment,enumerate,footnote,graphicx,subfloat,relsize}
\usepackage{array,tabularx,tabu,multirow,framed,afterpage}
\usepackage[usenames,dvipsnames]{xcolor}
\numberwithin{equation}{section}
\usepackage{cleveref}
\crefformat{pluralequation}{#2\black{eqs.~(}#1\black{)}#3}
\Crefformat{pluralequation}{#2\black{Equations~(}#1\black{)}#3}
\crefformat{pluralfigure}{#2\black{figs.~}#1#3}
\Crefformat{pluralfigure}{#2\black{Figures~}#1#3}

\usepackage[]{units}

\newcommand{\be}{\begin{equation}}
\newcommand{\ee}{\end{equation}}
\newcommand{\bee}{\begin{equation*}}
\newcommand{\eee}{\end{equation*}}
\newcommand{\bea}{\begin{eqnarray}}
\newcommand{\eea}{\end{eqnarray}}
\newcommand{\bean}{\begin{eqnarray*}}
\newcommand{\eean}{\end{eqnarray*}}

\newcommand{\lh}{\lambda_1}
\newcommand{\leta}{\lambda_2}
\newcommand{\lheta}{\lambda_3}
\newcommand{\muh}{\mu_1}
\newcommand{\mueta}{\mu_2}

\title{Strong gravitational radiation from a simple \\ dark matter model}

\author{Iason Baldes}
\emailAdd{iason.baldes@desy.de}
\author{and Camilo Garcia-Cely}
\emailAdd{camilo.garcia.cely@desy.de}

\affiliation{DESY, Notkestra{\ss}e 85, D-22607 Hamburg, Germany}

\date{November 29, 2018}

\arxivnumber{1809.01198}

\abstract{
A rather minimal possibility is that dark matter consists of the gauge bosons of a spontaneously broken symmetry. Here we explore the possibility of detecting the gravitational waves produced by the phase transition associated with such breaking. Concretely, we focus on the scenario based on an $SU(2)_D$ group and argue that it is a case study for the sensitivity of future gravitational wave observatories to phase transitions associated with dark matter. This is because there are few parameters and those fixing the relic density also determine the effective potential establishing the strength of the phase transition. Particularly promising for LISA and even the Einstein Telescope is the super-cool dark matter regime, with DM masses above $\mathcal{O}$(100) TeV, for which we find that the gravitational wave signal is notably strong. In our analysis, we include the effect of astrophysical foregrounds, which are often ignored in the context of phase transitions.
}

\begin{document}
\maketitle
\flushbottom

\section{Introduction}

Cosmological and astrophysical observations strongly suggest that, in contrast to the  ordinary substances found on Earth, baryons are not the dominant constituent of the matter in the Universe~\cite{Bertone:2004pz}. Such non-baryonic matter is called dark because its interactions with the Standard Model (SM) particles --- particularly with photons --- are constrained to be very weak. This, along with the obvious fact that dark matter (DM) must be stable on cosmological timescales, are the two most important properties of any DM candidate. 

The first property is often invoked as an argument for the electroweak (EW) nature of DM interactions. In fact, models where DM is directly coupled to the W or Z bosons
naturally explain the DM relic density by means of the thermal freeze-out of DM annihilations in the Early Universe. Nevertheless, these scenarios have been dramatically constrained in the past couple of decades by direct and indirect detection experiments, together with colliders, most recently the LHC~\cite{Kahlhoefer:2017dnp, Gaskins:2016cha, Undagoitia:2015gya}. 
In contrast, models where  DM is directly coupled to the Higgs and not to the W or Z bosons are much less constrained by the aforementioned experiments, especially in regimes where the DM is heavier than the Higgs boson. Interestingly, gravitational waves (GWs) offer a new complementary way  to probe the latter scenarios. This is because they typically require the existence of additional scalar fields, which can potentially trigger a first-order phase transition (PT) in the Early Universe and therefore the emission of GWs~\cite{Witten:1984rs,Hogan:1986qda,Grojean:2006bp,Caprini:2015zlo}.  Of course, DM can be probed in this way only if its properties are closely related to the PT~\cite{Schwaller:2015tja,Baldes:2017rcu,Chao:2017vrq,Croon:2018erz}. This is the subject of the present work.

In order to motivate a concrete choice for the DM model, we will invoke the second DM property mentioned above, i.e.~its stability. This is often ensured by imposing a discrete symmetry in the DM sector. The most common examples being $Z_2$ symmetries or the so-called R-parity in supersymmetric theories. Nevertheless, such symmetries are not known to exist in nature.\footnote{CPT is the only SM  discrete symmetry that is conserved.} Better-motivated scenarios are those where  DM is stable as a result of its own dynamics.  In fact, this is exactly what happens with the stable particles of the SM. For instance, proton stability follows from baryon  number conservation, which is an accidental symmetry  due to the $SU(3)_{C}\times SU(2)_{L} \times U(1)_{Y}$ charges of the matter fields.
An incomplete list of examples of this type of scenarios include Minimal DM~\cite{Cirelli:2005uq}, spin-one DM models~\cite{Hambye:2008bq,Lebedev:2011iq,Cata:2014sta,Arcadi:2016kmk} and QCD-like models of DM~\cite{Hochberg:2014kqa}.

The previous observations motivate us to study the spin-one DM model proposed in~\cite{Hambye:2008bq}, in which the DM portal to the SM is the Higgs boson.
Concretely, we extend the  SM with a dark $SU(2)_D$ local symmetry, under which all the SM particles are assumed to be singlets. In addition, we postulate a dark scalar doublet which carries no SM charges and whose vacuum expectation value (VEV) breaks the $SU(2)_D$ symmetry via a Higgs mechanism in the dark sector, ensuring the theoretical consistency of the model containing massive spin-one fields. 
After symmetry breaking,  the particle content includes --- besides the SM --- three mass-degenerate particles of spin-one and one dark Higgs boson. In this model there is a custodial $SO(3)$ symmetry remaining in the broken phase, under which the gauge bosons transform, ensuring their stability. Collectively, these comprise our DM candidate, which only couples to itself, to the SM Higgs $h$, and to the dark Higgs $h_D$. The Higgs portal interaction allows $h_{D}$ to decay to light SM particles, thus avoiding it becoming a DM component.

We will consider two production regimes for the DM relic density. First, the standard thermal freeze-out of DM annihilations into dark Higgs bosons.
Second, super-cool DM~\cite{Hambye:2018qjv}, a more exotic possibility in which we assume a classically scale invariant potential for our model~\cite{Hambye:2013sna,Carone:2013wla,Karam:2015jta,Chataignier:2018kay}. As pointed out recently, this can result in a period of late-time inflation which sets the relic density in a completely novel way. In both cases, a PT takes place in the early Universe from a $SU(2)_D$ symmetric vacuum in which the would-be-DM is massless, to a vacuum in which the dark gauge symmetry is broken and the DM is massive. The key point of our analysis is that the parameters setting the relic density also enter the effective potential determining the PT. As we will see, this allows us to find correlations between the GW signal and the DM properties. 

This study is timely, as much work is being done on understanding GWs from cosmological PTs in anticipation of LISA~\cite{Caprini:2015zlo}, and follow-up proposals such as BBO~\cite{Thrane:2013oya}. Our analysis differs from recent similar works in at least three aspects. First, as already mentioned, our scenario is rather minimal, with only four parameters in the general case and two for super-cool DM. This allows us to establish a close connection between the emission of  GWs and  the relic density or direct detection. Second,  in our analysis astrophysical foregrounds will be taken into account. These are mostly due to binaries of white dwarfs and are crucial for estimating the signal-to-noise ratio at future GW observatories~\cite{Hogan:1986qda,Grojean:2006bp}. Finally, we discuss for the first time the GW signatures of the super-cool DM regime. The paper is organized as follows. In Section~\ref{sec:model}, we present our DM model and its phenomenology. In Section~\ref{sec:GW}, we calculate the GW signal arising from the PT for the standard and the classically scale invariant cases. We conclude in Section~\ref{sec:conclusion} by presenting a summary and outlook for this work. Appendix~\ref{sec:effectivepotential} is devoted to details concerning the effective potential, which determines the nature of the PT, Appendix~\ref{sec:appB} summarises the contributions to the GW spectra, Appendix~\ref{sec:regb} includes some additional material regarding the classically scale invariant potential, and Appendix~\ref{sec:appC} discusses bubble percolation in the vacuum dominated regime.


\section{DM as massive gauge bosons} 
\label{sec:model}
\subsection{The model}

In this section we will describe the model and define notation. As mentioned in the introduction, we consider an extension of the SM with a dark $SU(2)_D$ gauge symmetry~\cite{Hambye:2008bq}, under which all the SM particles are singlets. In addition to the dark gauge bosons $A^i_{D\mu}$ $(i=1,2,3)$, the model has a dark scalar doublet, $H_D$, which carries no SM charges. Hence, the Lagrangian of the model is 
\begin{equation}
\label{Lag0}
{\cal L} = {\cal L}_\text{SM} -\frac{1}{4} F_D \cdot F_{D}+({\cal D} H_D)^\dagger ({\cal D} H_D) -\mu^2_2\,H_D^\dagger H_D-\leta\,(H_D^\dagger H_D)^2 -\lheta \,H_D^\dagger H_D\,H^\dagger H\,,
\end{equation}
where ${\cal L}_\text{SM} \supset -\muh^2 H^\dagger H- \lh(H^\dagger H)^2$ and $H$ is the SM scalar doublet. Here,  $F_D$  is the field strength tensor of the $SU(2)_D$ gauge symmetry and ${\cal D} =  \partial +i g_D \tau^{i} \cdot A_D^{i}/2 $ is the corresponding covariant derivative. 
We write scalar doublets as
\begin{align}
 H  = \frac{1}{\sqrt{2}} \begin{pmatrix}G^2+i G^3\\ \phi+ h+ i G^1\end{pmatrix}\;, &&H_D=\frac{1}{\sqrt{2}}\begin{pmatrix} G_D^2+i G_D^3\\  \eta +h_D +i G_D^1\end{pmatrix}\;,
\label{eq:fieldcomp}
\end{align}
where $\phi$ and $\eta$ are the classical field values breaking the EW and the $SU(2)_D$ symmetries, respectively. In addition, $h$, $h_D$, $G^i$ and $G^i_D$$ (i=1,2,3)$  are the corresponding Higgs and Goldstone boson fields.

\subsubsection*{Symmetry breaking at tree level}
In this case, the minimum of the potential associated with Eq.~\eqref{Lag0} is located at  $(\phi,\eta) = (v_\phi,v_\eta)$, where $v_{\phi}=\unit[246]{GeV}$ and 
	\begin{eqnarray}
	\muh^{2} =- \lh\,v_\phi^{2}-\frac{1}{2}\lheta \,v_\eta^{2}\,,\hspace{40pt}
	\mueta^{2} =- \leta v_\eta^{2}\,-\frac{1}{2}\lheta v_\phi^{2}\,.
	\label{eq:mass_parameters}
	\end{eqnarray}
The mixing of the real scalars is captured by the usual angle
	\begin{equation}
	\tan2\theta = \frac{\lheta \,v_\phi\,v_\eta}{\leta\,v_\eta^{2}\,-\,\lh\,v_\phi^{2}}\,.\label{kappa}
	\end{equation}
This is constrained by the Higgs signal strength measurements, $|\theta| \lesssim \mathcal{O}(0.1)$~\cite{Khachatryan:2016vau,Cheung:2018ave}, with the precise limit depending on which combination of measurements is taken.
For convenience we commit a small abuse of notation, and from now on also label the mass eigenstates with $h$ and $h_{D}$, where $m_{h} = \unit[125]{GeV}$. The mass eigenvalues are given by
	\begin{eqnarray}
	 m_{h}^{2} & = & 
 	2\,\lh\,v_\phi^{2}\,\cos^{2}\theta\,+\,2\,\leta\,v_\eta^{2}\,\sin^{2}\theta \label{mass1}
 	\,-\,\lheta v_\phi\,v_\eta\,\sin2\theta\,,\\
	 m_{h_D}^{2} & = & 
	2\,\lh\,v_\phi^{2}\,\sin^{2}\theta\,+\,2\,\leta\,v_\eta^{2}\,\cos^{2}\theta \label{mass2}
	 \,+\,\lheta v_\phi\,v_\eta\,\sin2\theta\,.
	\end{eqnarray}
All the dark gauge bosons obtain the mass, $m_A = g_D v_\eta/2$. In fact, they transform as a triplet under a
custodial $SO(3)$ symmetry. Notice the presence of light fermionic fields transforming under $SU(2)_D$ would spoil the stability of the vector DM, allowing the gauge bosons to decay, as occurs in the SM~\cite{Hambye:2008bq}. The absence of such fermions allows the model to remain rather minimal with only four parameters in the DM sector, which we take as $m_A$, $g_D$, $\theta$ and $m_{h_D}$.

\subsubsection*{Radiatively-induced symmetry breaking}

An alternative possibility is to consider a classically scale invariant realisation of this model~\cite{Hambye:2018qjv,Hambye:2013sna,Carone:2013wla,Karam:2015jta,Chataignier:2018kay}, where the mass terms in Eq.~(\ref{Lag0}) are forbidden and symmetry breaking is achieved through radiative effects. This is known as the Coleman-Weinberg mechanism~\cite{Coleman:1973jx,Gildener:1976ih}. A systematic analysis of radiative symmetry breaking with the above field content can be found, e.g.~in~\cite{Chataignier:2018kay}. In the present analysis,
the parameter regime of interest corresponds to what has been termed sequential symmetry breaking~\cite{Chataignier:2018kay}. The running of $\lambda_{2}$ results in it turning negative in the IR, breaking the $SU(2)_D$ symmetry via the Coleman-Weinberg mechanism.\footnote{The $\beta$ functions can be found in~\cite{Hambye:2013sna}.} 
If $\lambda_{3}<0$, the breaking of EW symmetry follows sequentially from the induced tachyonic mass $\lambda_3v_\eta^{2}/4$, which leads to $v_\phi = v_\eta\sqrt{-\lambda_3/(2\lambda_1)}$. Since we are interested in DM above the EW scale, i.e. $v_\phi \ll v_\eta$, the magnitude of the portal  must be  very small, $|\lambda_{3}| \ll 1$. This implies that the approximation of ignoring the $\phi$ direction in studying the $SU(2)_D$ symmetry breaking is consistent. Under these assumptions, we can study $SU(2)_D$ symmetry breaking by focusing on the term  ${\cal L} \supset -\lambda_2\eta^4/4$, where the coupling $\lambda_2$ is evaluated at a sliding scale given by the  value of the $\eta$ field, giving 
\begin{equation}
\lambda_2(\eta)\approx \frac{9 \,g_D^4}{128 \pi^2} \mathrm{Ln} \left(\frac{\eta}{\eta_0}\right)\,,
\label{eq:l2}
\end{equation}
with $\eta_0$ being the scale at which $\lambda_2$ flips sign.   Here, we neglect the contributions of $\lambda_2$ and $\lambda_3$ to R.H.S. of Eq.~(\ref{eq:l2}), which is a valid approximation provided $\lambda_{2} \ll g_{D}^{4}$ and $\lambda_3^{2} \ll g_{D}^{4}$~\cite{Chataignier:2018kay}. We also ignore the running of $g_{D}$. Note the dark and visible sectors are close to decoupled not only because the portal coupling is small but also because the corresponding beta function is proportional to $\lambda_3$. In fact,  the running of the latter between $v_{\eta}$ and $v_{\phi}$ is not so large as to affect our analysis.\footnote{One can make this statement more precise by  considering a scalar potential improved  with  Renormalization-Group effects, as recently suggested in Refs.~\cite{Chataignier:2018aud, Chataignier:2018kay}. This can cure any potential pitfall associated to the disparity of VEVs in the scalar potential. However, such analysis lies beyond the scope of this work.}

As alluded to above, $\lambda_{2}<0$ signals the breaking of the $SU(2)_D$ gauge symmetry.
In fact,  the minimization conditions  to leading order in $\lambda_3$, give $v_\eta = \eta_0 e^{-1/4}$ together with 
\begin{align}
&& m_{h_D}^2= \frac{9 \,g_D^4}{128 \pi^2} v_\eta^2&&\text{and}&& m_h^2 = -\lambda_3 v_\eta^2 \,.
\label{eq:RISB}
\end{align}
As in the previous case, the dark gauge bosons obtain a mass $m_A = g_D v_\eta/2$. Notice also that, after accounting for $m_{h} = \unit[125]{GeV}$ and $v_{\phi}=\unit[246]{GeV}$, there are only two free parameters, which we choose as $m_A$ and $g_D$. 
Before discussing DM production, we would like to emphasize that this scenario is not simply a limit of the previous case when $\mu_1$ and $\mu_2$ approach zero because here the breaking of the symmetry does not occur at tree level. (For a detailed discussion on such a limit, see~\cite{Coleman:1973jx}.)

\subsection{Relic density}

\begin{figure}[t]
\begin{center}
\includegraphics[width=440pt]{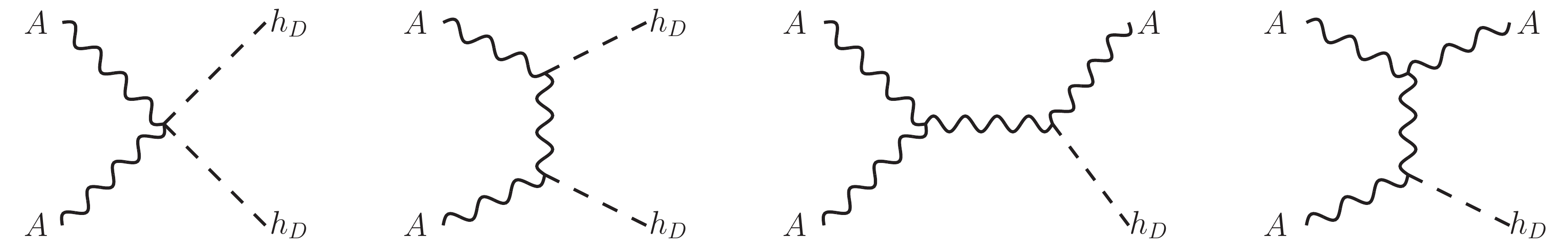}
\end{center}
\caption{The dominant DM annihilation channels for $m_{A} \gg m_{h_D}$ and $\theta \ll 1$.}
\label{fig:freezeout}
\end{figure}

We will consider two production regimes for the DM relic density: the standard freeze-out scenario and super-cool DM.\footnote{Other production mechanism for this model have been discussed in~\cite{Bernal:2015ova}.} The latter only takes place for the classically scale invariant case, i.e. when the gauge symmetry is broken radiatively. Details are given in section \ref{sec:scaleinv}. For the former case, we make the mild assumption that $\lambda_{3}$ and $g_{D}$ are large enough so that DM was in thermal equilibrium with the SM fields in the Early Universe. Freeze-out leads to the observed dark matter abundance, $\Omega h^2 \simeq 0.12$, when the corresponding cross section is of the order $\unit[2.3 \times 10^{-26}]{cm^3/s}$. This means that for given $m_A$, $m_{h_D}$, and $\theta$, the relic density fixes the dark coupling $g_D$. We are interested in the regime in which $m_A > 2m_{h_D}$ so that DM (semi-)annihilates into dark Higgs bosons. We make the further simplifying assumption, $m_{A} \gg m_{h_D}$ and $\theta \ll 1$, so that the annihilations into SM particles by means of a scalar exchange in the s-channel are negligible and the dominant  annihilation channels are those shown in Fig.~\ref{fig:freezeout}. In this regime the correct relic density is achieved for,
	\begin{equation}
	\label{eq:relic}
	g_{D} \approx 0.9 \times \sqrt{\frac{m_{A}}{1 \; \mathrm{TeV}} }\hspace{10pt} \text{and} \hspace{10pt}v_\eta \approx \unit[2.2]{TeV} \times \sqrt{\frac{m_{A}}{1 \; \mathrm{TeV}} }.
	\end{equation}
A more accurate determination can be achieved by numerically solving the Boltzmann equations. Given the uncertainties of the gravitational wave spectrum, however, the use of Eq.~(\ref{eq:relic}) is sufficient for our purposes. The coupling $g_{D}$ is fixed by the relic abundance, which effectively collapses the higher dimensional parameter space to three (one) dimensions in the standard (classically scale invariant) case, which would otherwise have to be scanned over.

\subsection{Direct detection}

\begin{figure}[t]
\begin{center}
\includegraphics[width=210pt]{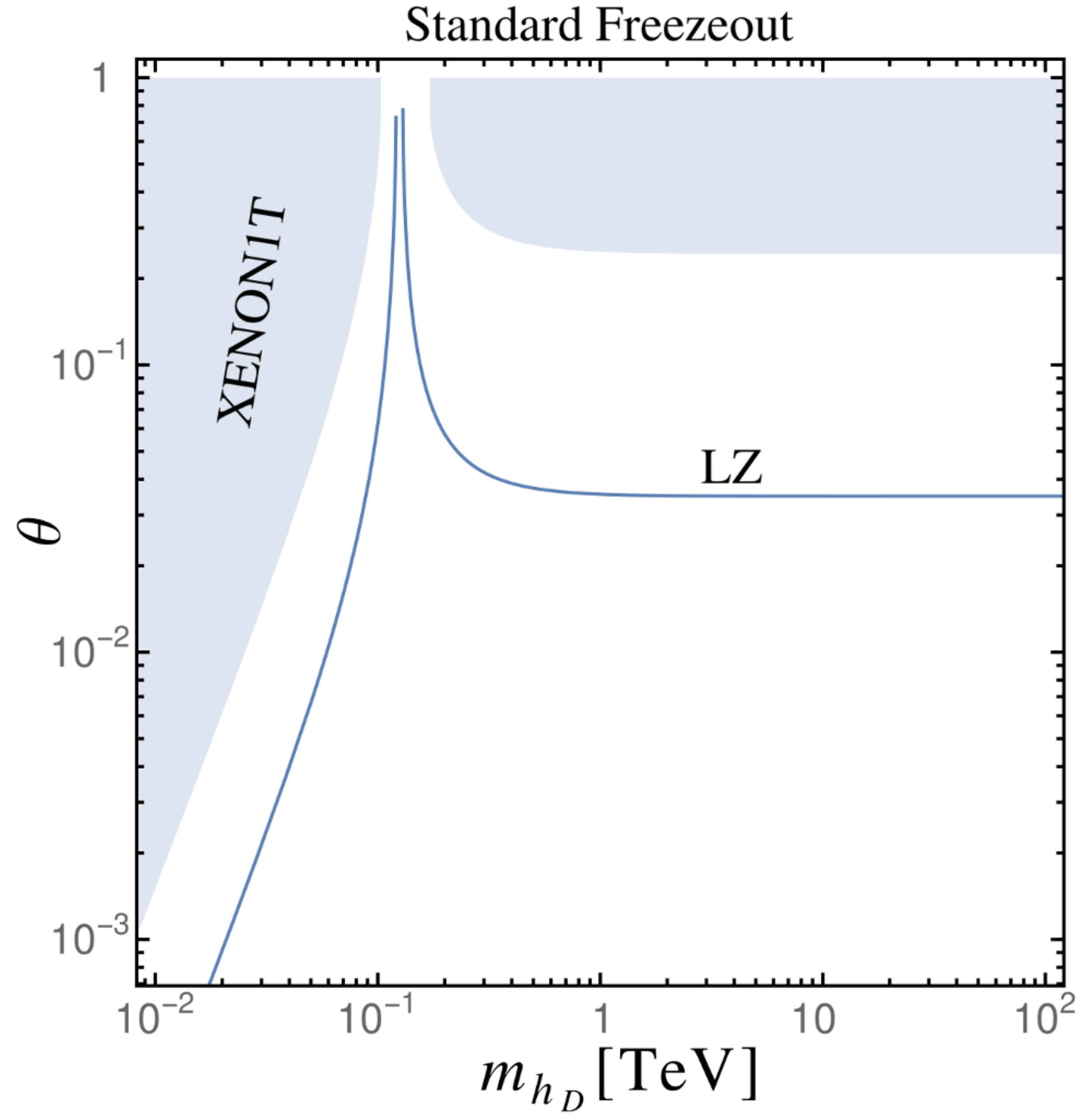}
\includegraphics[width=210pt]{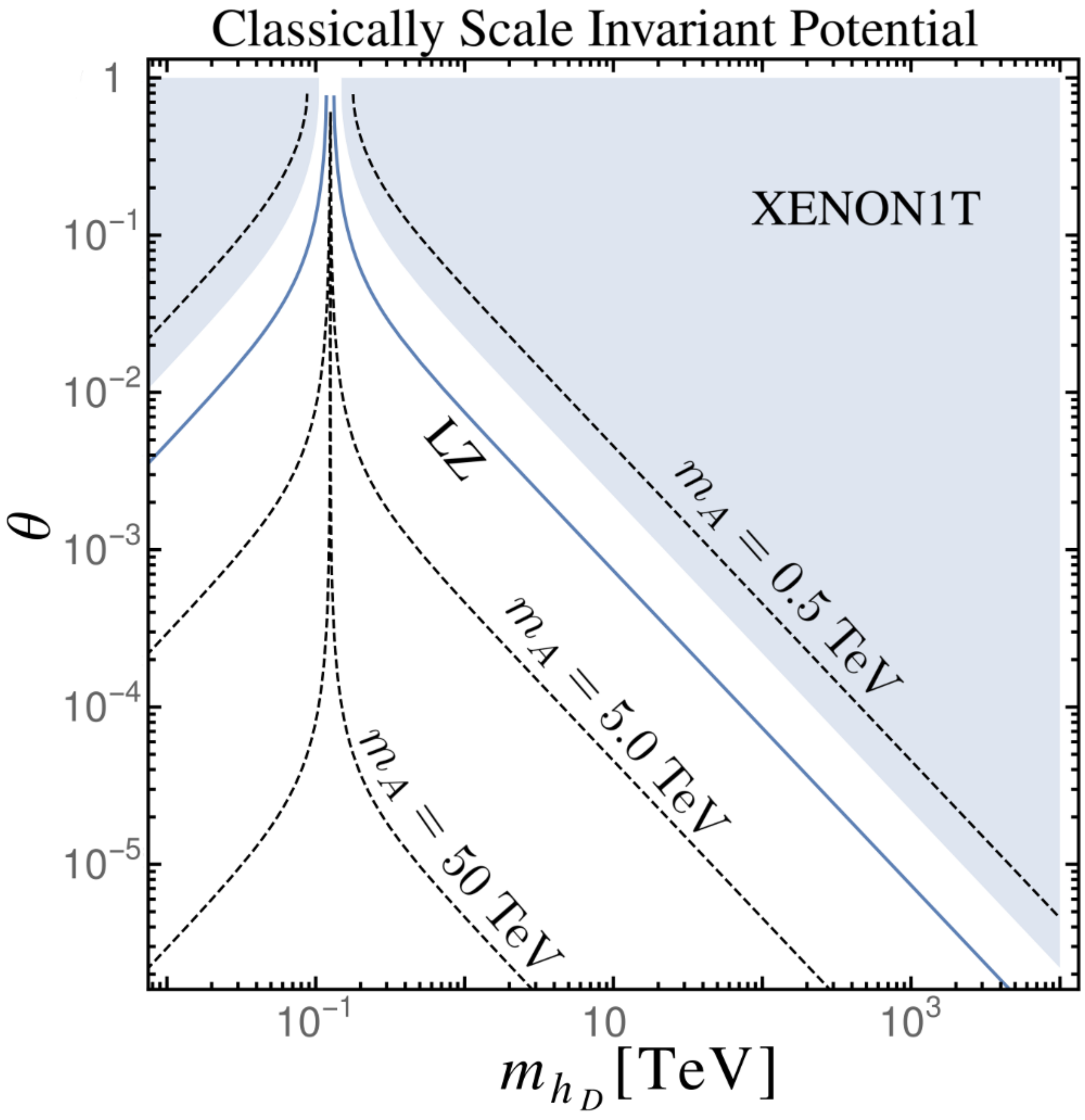}
\end{center}
\caption{Mixing angles excluded by Xenon1T~\cite{Aprile:2017iyp} (shaded area)  together with the projected sensitivity from LZ~\cite{Mount:2017qzi} (solid line). Here we assume $m_A\gtrsim \unit[0.1]{TeV}$. \emph{Left:} assuming the relic density is produced via the freeze-out mechanism. \emph{Right:} for the classically scale invariant potential, including contours for some choices of the DM mass.}
\label{fig:DD}
\end{figure}

The spin independent scattering cross-section of dark matter off nucleons is~\cite{Hambye:2008bq}
\begin{equation}
\sigma_{\rm SI} =
\frac{ g_{D}^{4} \, f^{2}  \, m_{N}^{4}  \, v_{\eta}^{2}   }{ 64 \pi \, (m_{N} + m_{A} )^{2} \, v_{\phi}^{2}  } \left( \frac{1}{m_{h}^{2}} - \frac{1}{m_{h_{D}}^{2}}\right)^{2} \, \sin^{2}{2\theta}, 
\label{DirectDetectionEq}
\end{equation}
where $m_N$ denotes the nucleon mass and $f \simeq 0.3$ is a constant that depends on the nucleon matrix element. Thus, current experiments such as Xenon1T~\cite{Aprile:2017iyp,Cui:2017nnn,Akerib:2016vxi} constrain the mixing angle between the scalars. This is shown in Fig.~\ref{fig:DD} assuming $m_A \gtrsim \unit[0.1]{TeV}$ together with the corresponding future sensitivity from LUX-Zepelin~\cite{Aprile:2015uzo,Akerib:2015cja,Mount:2017qzi,Aalbers:2016jon}. In the left panel, the gauge coupling is fixed by freeze out. In the right panel, we consider radiatively-induced symmetry breaking and therefore we use Eq.~\eqref{eq:RISB} to calculate the cross section. No production mechanism is assumed in the latter case.
 
The left panel of the figure can be understood as follows. Working in the limit $m_{A} \gg m_{h_{D}} \gg m_{h}$ and substituting Eq.~(\ref{eq:relic}) into $\sigma_{\rm SI}$, one finds $\sigma_{\rm SI} \propto m_{A} \sin^{2}{2\theta}$, i.e.~with no further dependence on the other DM parameters. Then comparing to direct detection constraints --- which also scale as $\sigma_{\rm SI}^{\rm limit} \propto m_{\rm DM}$ for  $m_{\rm DM} \gtrsim 100$ GeV --- we find current experiments such as Xenon1T demand $\theta \lesssim \mathcal{O}(0.1)$. Future experiments such as LUX-Zepelin, will improve $\sigma_{\rm SI}^{\rm limit}$ by two orders of magnitude and therefore probe down to $\theta \sim \mathcal{O}(0.01)$.
On the other panel, Eq.~\eqref{eq:RISB} indicates that $\sigma_{\rm SI} \propto 1/m_{A}^6 $ with no further dependence on the other dark sector parameters. Hence, $\sigma_{\rm SI}^{\rm limit} \propto m_{\rm DM}$ excludes contours of constant DM mass.  In fact, for the classically invariant case, the  current direct detection constraint demands $m_{A} \gtrsim 0.9$ TeV~\cite{Hambye:2018qjv}.


\section{Gravitational waves}
\label{sec:GW}
\subsection{Calculation of the spectrum}

A first-order PT takes place by means of the nucleation of true-vacuum bubbles in a false-vacuum background. At a given temperature $T$, the rate per unit of volume at which this occurs scales as $T^4 e^{-S}$, where  $S$ is the Euclidean action evaluated at the solution describing one bubble.  In practice, $S$ can be approximated by the smallest of $S_3/T$ or $S_4$,  where $S_n$   is the $O(n)$ symmetric action.
For our model,  we have checked\footnote{More precisely, after calculating the nucleation temperature by means of Eq.~\eqref{eq:comparisionSvsH}, we also computed $S_4$ for the parameter points which exhibit the largest supercooling (including, of course, points in the vacuum dominated regime). For all the points we checked, we found $S_3/T < S_4$. } that $ S_3/T < S_4 $  and therefore
	\begin{equation}
	S\approx \frac{S_3}{T} =\frac{4\pi}{T}\int r^{2} \left\{\frac{1}{2}\left(\frac{d\phi}{dr}\right)^{2}+\frac{1}{2}\left(\frac{d\eta}{dr}\right)^{2}+V(\phi,\eta,T)-V(\phi_0,\eta_0,T)\right\}dr.
	\label{eq:S3}
	\end{equation}

Here,  $r$ is the radial coordinate of the bubble, $V(\phi,\eta,T)$ is the finite-temperature effective potential, which we calculate using the well-known techniques of thermal field theory, and $(\phi_0,\eta_0)$ are the field values of the false vacuum. The thermal functions which enter $V(\phi,\eta,T)$ are evaluated numerically. Further details are given in Appendix~\ref{sec:effectivepotential}. To calculate $S_{3}$, we use the over/under shooting method implemented in our code to find the bubble profile by solving the equations of motion,
	\begin{equation}
	\frac{d^{2}\phi}{dr^{2}}+\frac{2}{r}\frac{d\phi}{dr}=\frac{\partial V}{\partial \phi}, \qquad  \qquad  \qquad \frac{d^{2}\eta}{dr^{2}}+\frac{2}{r}\frac{d\eta}{dr}=\frac{\partial V}{\partial \eta},
	\end{equation}
with the boundary conditions $d\phi/dr\big|_{r=0}= d\eta/dr\big|_{r=0}=0$ and $\phi|_{r\to\infty}=\phi_0$ and $\eta|_{r\to\infty}=\eta_0$. Physically, these conditions correspond to demanding a smooth profile at the centre of the bubble, $r=0$, together with the Universe being in the false vacuum well outside of the bubble, $r \to \infty$. In our discussion here we have remained general by including both fields, however, in the PTs studied below only the $\eta$ field value will be changing which simplifies our calculation of the action.

Nucleation occurs at a temperature $T_{n}$, when the bubble nucleation rate in the horizon volume becomes comparable to the Hubble parameter, from now on denoted $H$.
Hence, we find the nucleation temperature by solving~\cite{Moreno:1998bq}
	\begin{align}
	\frac{S_{3}}{T} & \approx 4 \, \mathrm{Ln}\left(  \frac{ T}{H}\right) \label{eq:nucleation}  \\
			& \approx 
		 \begin{dcases*}
		146 - 4 \; \mathrm{Ln}\left( \frac{T}{100 \; \mathrm{GeV} } \right) - 2 \; \mathrm{Ln}\bigg(\frac{g_{\ast}}{100}\bigg), & radiation dominated, \\
		135 + 4 \; \mathrm{Ln}\left( \frac{T}{100 \; \mathrm{GeV} } \right) - 8 \; \mathrm{Ln}\left(\frac{\rho_{\rm vac}^{1/4}}{1 \; \mathrm{TeV} }\right), & vacuum dominated,
		\end{dcases*}
\label{eq:comparisionSvsH}
	\end{align}
where $\rho_{\rm vac}$ is the vacuum energy density, and $g_{\ast}$ counts the effective radiation degrees of freedom.
With the field content of the present model, $g_{\ast}=116.75$ when all species are relativistic and thermalised.  For the parameter regions of interest in this work, we find that the nucleation temperature is significantly different from the one associated with EW symmetry breaking.  As a result, the phase transition --- and therefore the GW emission --- is mostly determined by the one-dimensional $\eta$ direction.

An isolated spherical bubble does not radiate gravitationally because its quadrapole moment is zero. In contrast, the collision of several bubbles generates GWs by means of at least three different processes:  collision of bubble walls (mostly determined by the dynamics of the scalar fields), bubble percolation producing sound waves, and magnetohydrodynamic turbulence in the plasma. These give rise to 
\begin{equation}
	h^{2}\Omega_{\rm GW}(f) \equiv h^{2}\frac{f}{\rho_{c}}\frac{d  \rho_{\rm GW}}{d f},
	\end{equation}
where $\rho_{c}$ is the critical density, and $d \rho_{GW}/df$ is the differential GW energy density.  Its determination is an active area of research with a large number of ongoing investigations. This is illustrated by the fact that the study of GWs from spin-one DM that was briefly discussed in Ref.~\cite{Hambye:2013sna} only accounted for the collision of bubble walls. Nevertheless, as discussed at length below, recent developments~\cite{Hindmarsh:2013xza} when applied to our model indicate that sounds waves and turbulence give the dominant contribution, at least for the case when symmetry breaking occurs at tree level.
 
In the light of this, here we use the compendium of results presented in Ref.~\cite{Caprini:2015zlo} and summarized in Appendix~\ref{sec:appB} in finding the GW spectrum. The spectrum depends on four parameters: the Hubble parameter at nucleation, the wall velocity, $v_w$, the latent heat (here normalised to the radiation density $\rho_{\rm rad}$),
	\begin{equation}
	\alpha  = \frac{1}{\rho_{\rm rad}} \bigg( 1 -  T\frac{\partial}{\partial T} \bigg) \bigg (V [\phi_0,\eta_0] - V [\phi_n,\eta_n] \bigg) \bigg|_{T_n}
	\label{eq:alpha}
	\end{equation}
where  $(\phi_n,\eta_n)$ is the true vacuum at $T_n$, and the timescale of the transition, 
	\begin{equation}
       \beta    = H\, T_{n} \frac{d}{dT}\left(\frac{S_3}{T}\right)\bigg|_{T_n}\,.
	\label{eq:beta}
	\end{equation}
For the strongest of PTs, we expect the wall velocity to be close to luminal, $v_{\rm w} \simeq 1$. This is because the mean field potential typically satisfies,
 	\begin{align}
	\overline{V}  = & \; V(\phi_n,\eta_n,T=0) -V(\phi_0,\eta_0,T=0) \\
			& \; + \frac{T^{2}}{24}\left(\sum_{\rm bosons}\left[ m_{b}^{2}(\phi_n,\eta_n)-m_{b}^{2}(\phi_0,\eta_0) \right]+\frac{1}{2}\sum_{\rm fermions}\left[m_{f}^{2}(\phi_n,\eta_n)-m_{f}^{2}(\phi_0,\eta_0) \right] \right) < 0. \nonumber
	\end{align}
This is the Bodeker-Moore (BM) criterion~\cite{Bodeker:2009qy}. We shall make clear on our plots where the BM criterion holds and what we assume regarding $v_w$ when it does not.

\begin{figure}[t]
\includegraphics[width=0.55\textwidth]{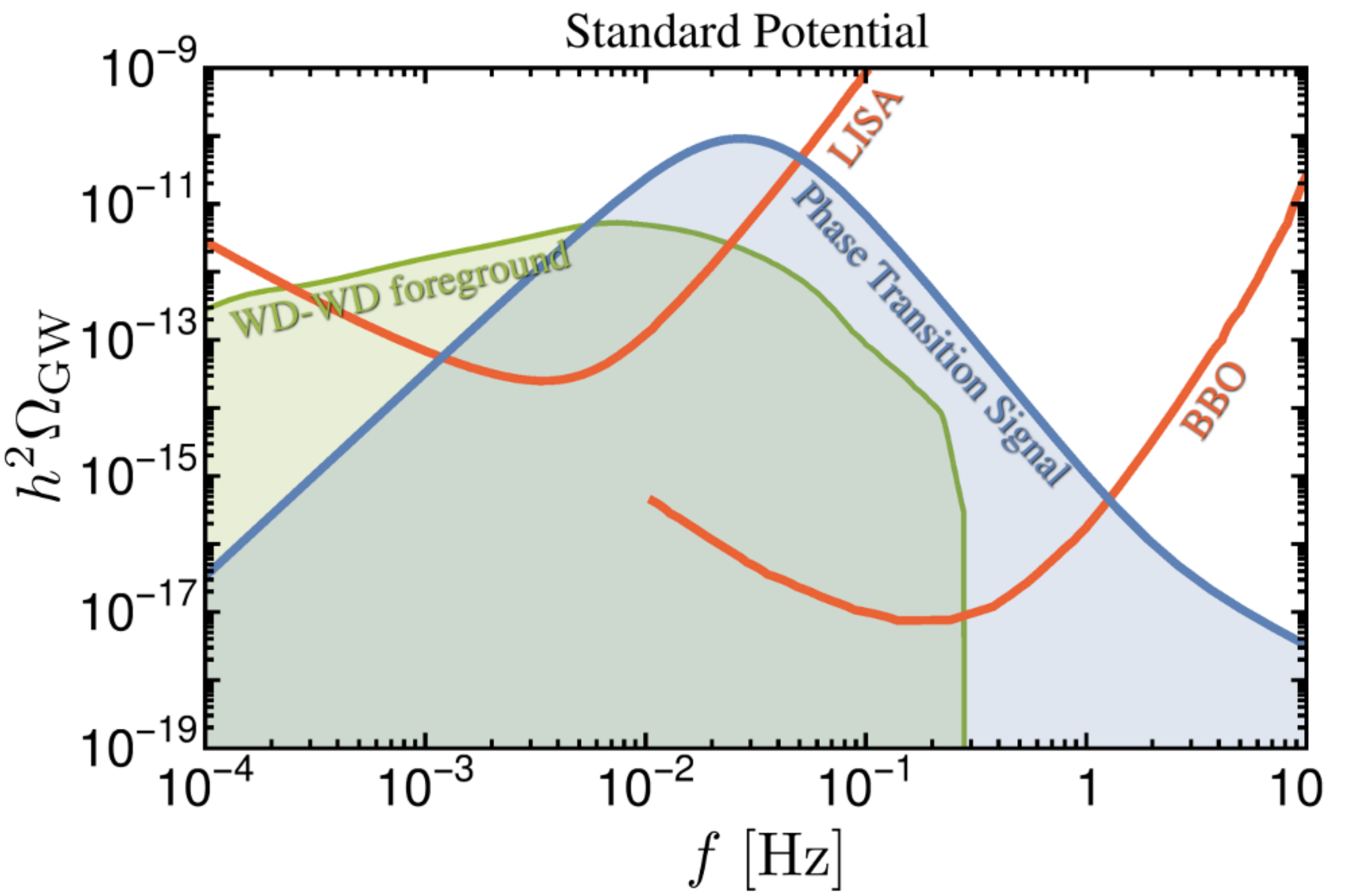}
   \hspace{30pt}
\raisebox{0.15\height}{\includegraphics[width=0.35\textwidth]{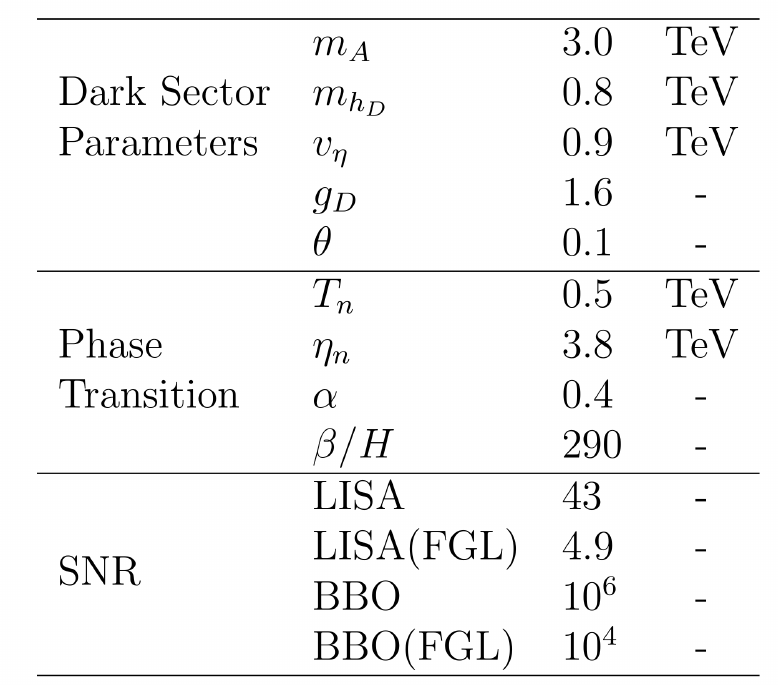}}
\caption{An example of the gravitational wave spectrum when the symmetry breaking occurs at tree level, together with the white-dwarf white-dwarf binary foreground, and LISA and BBO sensitivity curves.  
Here we assume $v_{w}=1$. 
}
\label{fig:spectrum}
\end{figure}

Even when the BM criterion holds, however, the wall is not expected to runaway with $\gamma \to \infty$, due to the transition radiation effect from the gauge bosons~\cite{Bodeker:2017cim}.
Therefore, provided the gauge boson population has not been overly diluted by false vacuum inflation, energy released in the transition is transfered to the radiation bath, in which sound waves~\cite{Hindmarsh:2013xza,Hindmarsh:2015qta,Hindmarsh:2017gnf} and magnetohydrodynamic turbulence~\cite{Caprini:2009yp,Binetruy:2012ze}, rather than the bubble wall collisions directly~\cite{Kosowsky:1991ua,Kosowsky:1992rz,Kosowsky:1992vn,Kamionkowski:1993fg,Caprini:2007xq,Huber:2008hg}, lead to a gravitational wave signal.

In summary, determining $T_{n}$, $\alpha$, and $\beta$ from Eqs.~\eqref{eq:S3}, \eqref{eq:nucleation}, \eqref{eq:alpha} and \eqref{eq:beta} allows us to find  $h^{2}\Omega_{\rm GW}(f)$ by means of the spectra summarized in Appendix \ref{sec:appB}. An example of the spectrum is shown in Fig.~\ref{fig:spectrum}. We use estimated sensitivity of the gravitational wave detectors to stochastic backgrounds, $h^{2}\Omega_{\rm sens}(f)$, for LISA~\cite{Caprini:2015zlo}, BBO~\cite{Thrane:2013oya}, and when applicable the Einstein Telescope (ET)~\cite{Punturo:2010zz,Hild:2010id,Regimbau:2011rp} (for which we use the updated sensitivity curve from~\cite{Cui:2018rwi}). The signal-to-noise ratio can be estimated using~\cite{Caprini:2015zlo}
	\begin{equation}
	\mathrm{SNR} = \sqrt{t_{\rm obs} \int \left[\frac{h^{2}\Omega_{\rm GW}(f)}{h^{2}\Omega_{\rm sens}(f)}\right]^{2} df}, 
	\end{equation}
where $t_{\rm obs}$ is the time of observation in years. We assume $t_{\rm obs} = 5$ throughout.

Confusion noise from astrophysical foregrounds may be an issue at these frequencies. We shall compare to some estimates of the unresolvable components given in the literature. The ensemble of white dwarf - white dwarf (WD-WD) binaries are thought to be the dominant source of this foreground, exceeding the unresolvable neutron star - neutron star (NS-NS) foreground~\cite{Farmer:2003pa,Rosado:2011kv}. In this work we restrict ourselves to the foreground from the extragalactic WD-WD ensemble and use the central value given in~\cite{Farmer:2003pa}. We make this choice because, in contrast to the extragalactic ensemble, it is thought the WD-WD Galactic foreground~\cite{Kosenko:1998mv,Nelemans:2001hp,Ruiter:2007xx} can be subtracted~\cite{Adams:2010vc,Adams:2013qma}. The continuous extragalactic NS-NS foreground extends to higher frequencies in the BBO band, however, it is thought that this can also be subtracted~\cite{Cutler:2005qq,Abbott:2017xzg}. We also adopt an alternative, foreground-limited, estimate of the signal-to-noise ratio
	\begin{equation}
	\mathrm{SNR}_{\rm FGL} = \sqrt{t_{\rm obs} \int \left[\frac{h^{2}\Omega_{\rm GW}(f)}{h^{2}\Omega_{\rm sens}(f)+h^{2}\Omega_{\rm FG}(f)}\right]^{2} df}, 
	\label{eq:snrfgl}
	\end{equation}
in which we attempt to naively capture the degradation of the sensitivity once the foreground, $h^{2}\Omega_{\rm FG}(f)$, is taken into account. The aim of introducing Eq.~(\ref{eq:snrfgl}) is to be able to roughly capture, in a single number, whether the signal extends above the sensitivity and foreground estimate. Whether such a PT signal could actually be separated from the astrophysical foreground depends, of course, on a myriad of factors, e.g. the robustness of the estimates of the amplitudes and spectral shapes of the signal and foreground, together with the confidence in our knowledge of the instrumental noise. These are topics worthy of further study, but we will not attempt to do them justice here. Nevertheless, we would like to remark that the LISA $\mathrm{SNR}_{\rm FGL}$ value  associated to the spectrum of  Fig.~\ref{fig:spectrum}  clearly illustrates the importance of astrophysical foregrounds, even though they are often ignored in similar studies. Furthermore, we wish to emphasize that our sensitivity analysis in terms of SNR significantly improves that from Ref.~\cite{Hambye:2013sna}, where GWs from spin-one DM were briefly discussed.

\subsection{Symmetry breaking at tree level}
\label{sec:standard}

\begin{figure}[t]
\begin{center}
\includegraphics[width=200pt]{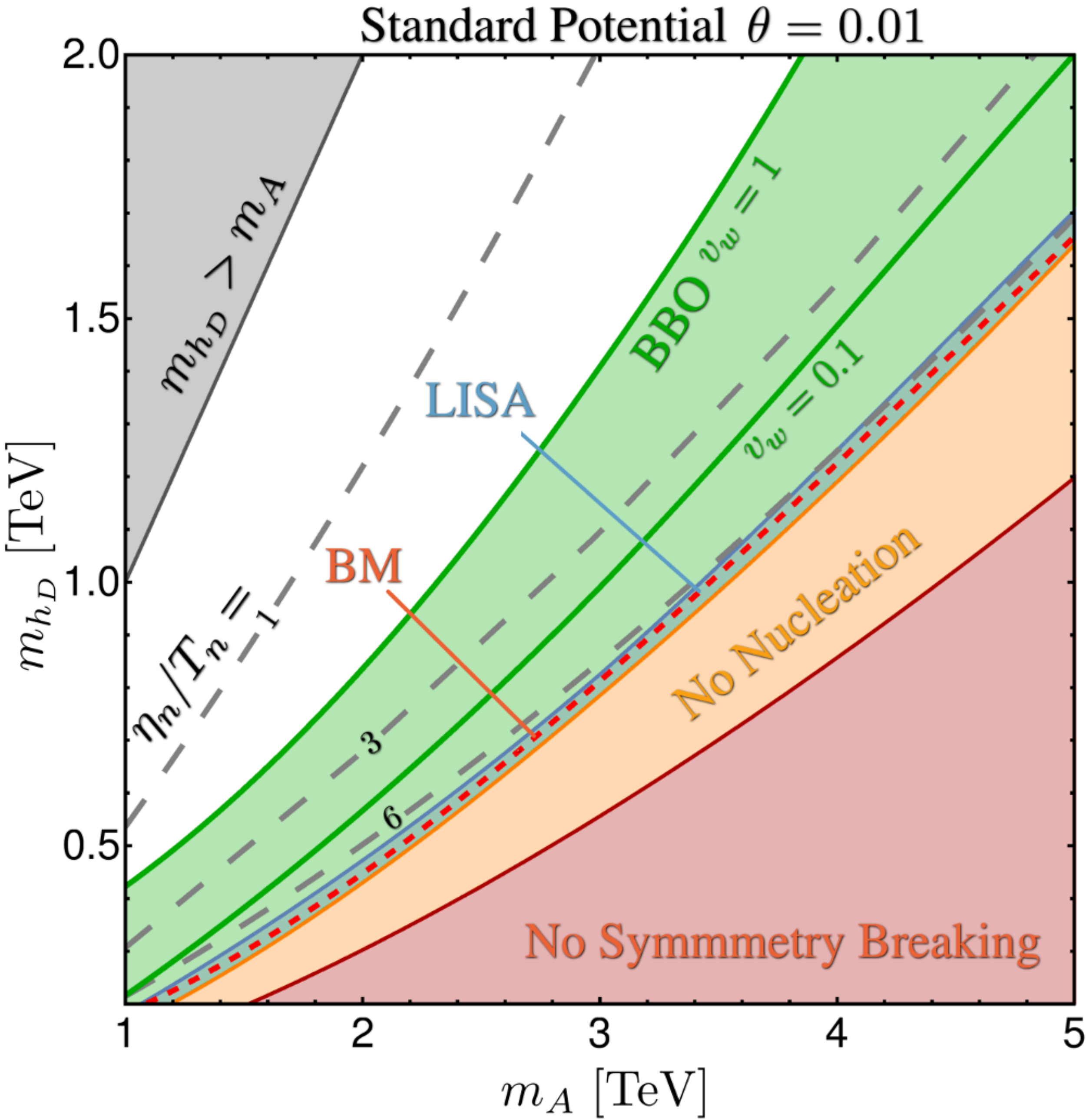}
\includegraphics[width=200pt]{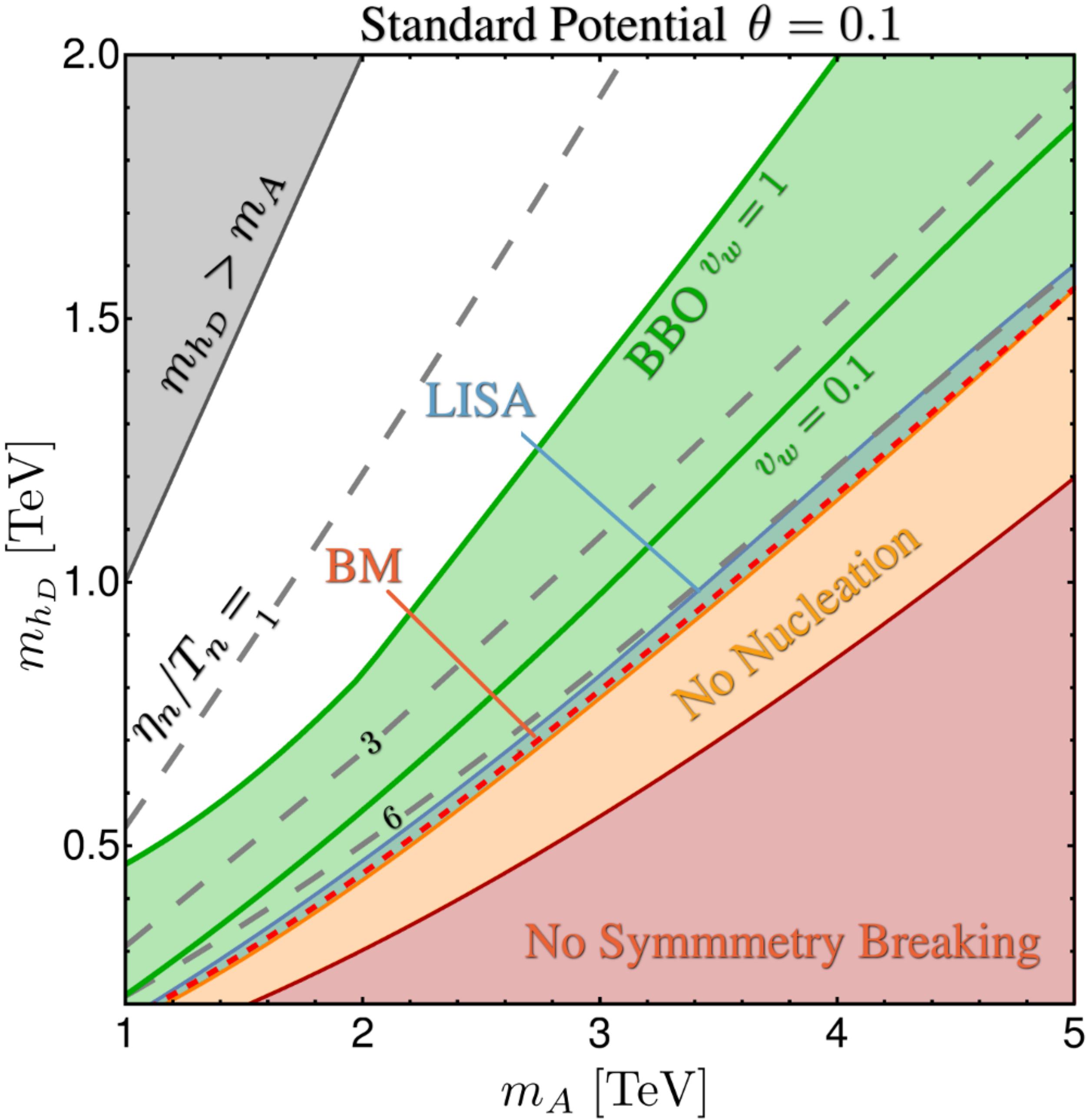}
\end{center}
\caption{The parameter space returning a significant BBO or LISA signal, SNR $>5$, when the symmetry breaking occurs at tree level (standard potential). For LISA we assume $v_{w}=1$ as the BM criterion is fulfilled roughly in this region. For BBO we show contours assuming  $v_{w} = 0.1$ and $1$. Only the strongest transitions, close to the point at which no transition occurs at all, can be probed by LISA in this case. In contrast BBO can probe a substantial fraction of the parameter space with a strong first order phase transition. Here we show the SNR with no foreground. If the foreground is included the BBO area remains practically unchanged, while the already small LISA area is approximately halved.}
\label{fig:largevev}
\end{figure}

\begin{figure}[t]
\begin{center}
\includegraphics[width=200pt]{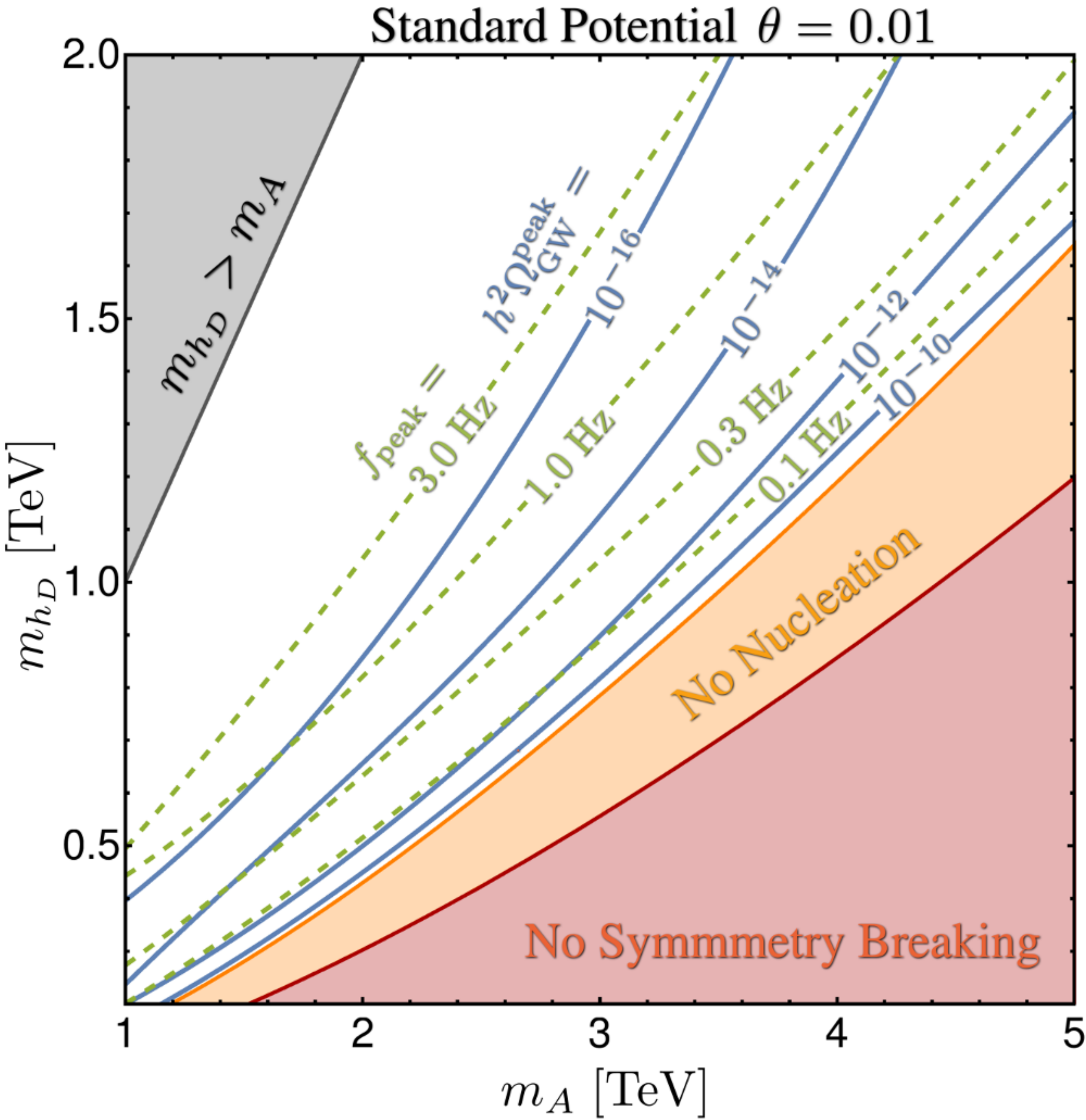}
\includegraphics[width=200pt]{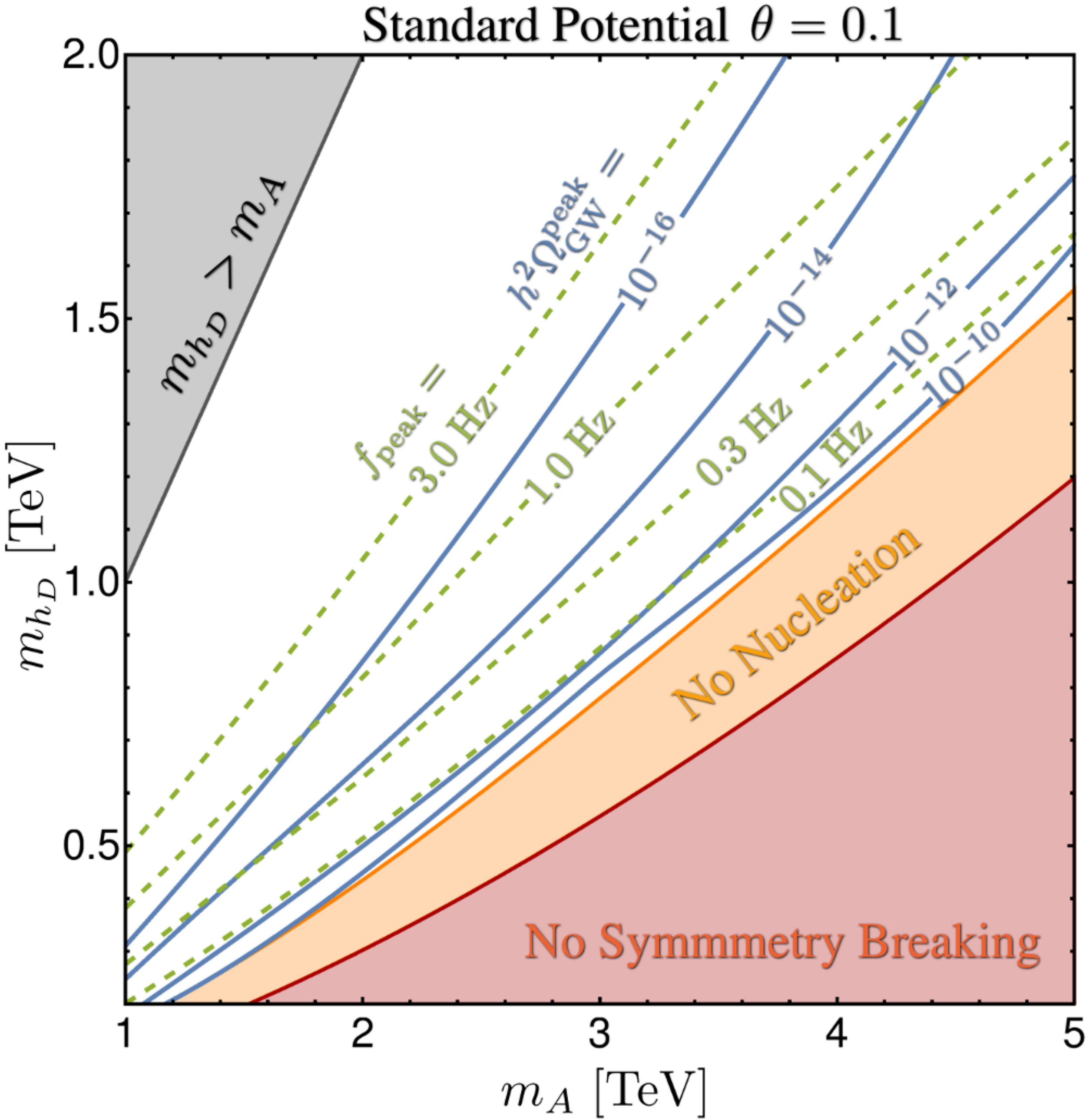}
\end{center}
\caption{Similar to Fig.~\ref{fig:largevev}, but with contours of the peak frequency and the corresponding GW energy density assuming $v_{w}=1$.}
\label{fig:largevev2}
\end{figure}

As discussed above, in this case the DM production proceeds via the standard freeze-out mechanism. Interesting for us is the regime with a large $g_{D}$, as this will lead to a strong phase transition. This pushes us to large $m_{A}$ and $v_{\eta}$, see Eq.~\eqref{eq:relic}, and  the dark phase transition will generally occur prior to the EW one. Thus the task of studying the phase transition reduces to one dimension in field space. We have seen an example of the gravitational wave spectrum, together with the dominant foreground, in Fig.~\ref{fig:spectrum}. We have also fixed $\theta$ to 0.1 and 0.01 (motivated by present and future direct detection constraints as shown in the left panel of Fig.~\ref{fig:DD}), scanned over the parameters $m_{A}$ and $m_{h_D}$,\footnote{By manually choosing the points, we were able to obtain acceptable fits for the various contours with the PT parameters calculated for $\sim$ 100 points in total.} and calculated the GW signal. The result is shown in Fig.~\ref{fig:largevev}. Likewise, using the expressions of Appendix~\ref{sec:appB}, we calculate the peak frequency and the peak GW energy density for each point of the parameter space and show the results in Fig.~\ref{fig:largevev2}.

These plots can be understood as follows. A larger DM mass, $m_{A}$, requires a larger gauge coupling in order to return the observed DM density. This results in a stronger phase transition from the one-loop effects of the gauge bosons. Similarly, in analogy with the SM, a lighter dark Higgs --- corresponding to a smaller quartic $\lambda_{2}$ ---  also leads to a stronger transition because the broken phase minimum is shallower. Nevertheless, for particularly large values of $m_{A}/m_{h_D}$, the one-loop effects can raise the broken phase minimum too far, resulting in the Universe becoming stuck in the symmetric phase. The latter can either be a false or true minimum, corresponding to the orange and red shaded regions of the figures respectively. We expect the allowed parameter space to be increased somewhat, into the orange region, if $S_{4}$ nucleation at lower temperatures were to be taken into account.

\subsection{Radiatively-induced symmetry breaking: standard freeze-out and super-cool DM}
\label{sec:scaleinv}

\begin{figure}[t]
\begin{center}
\includegraphics[width=200pt]{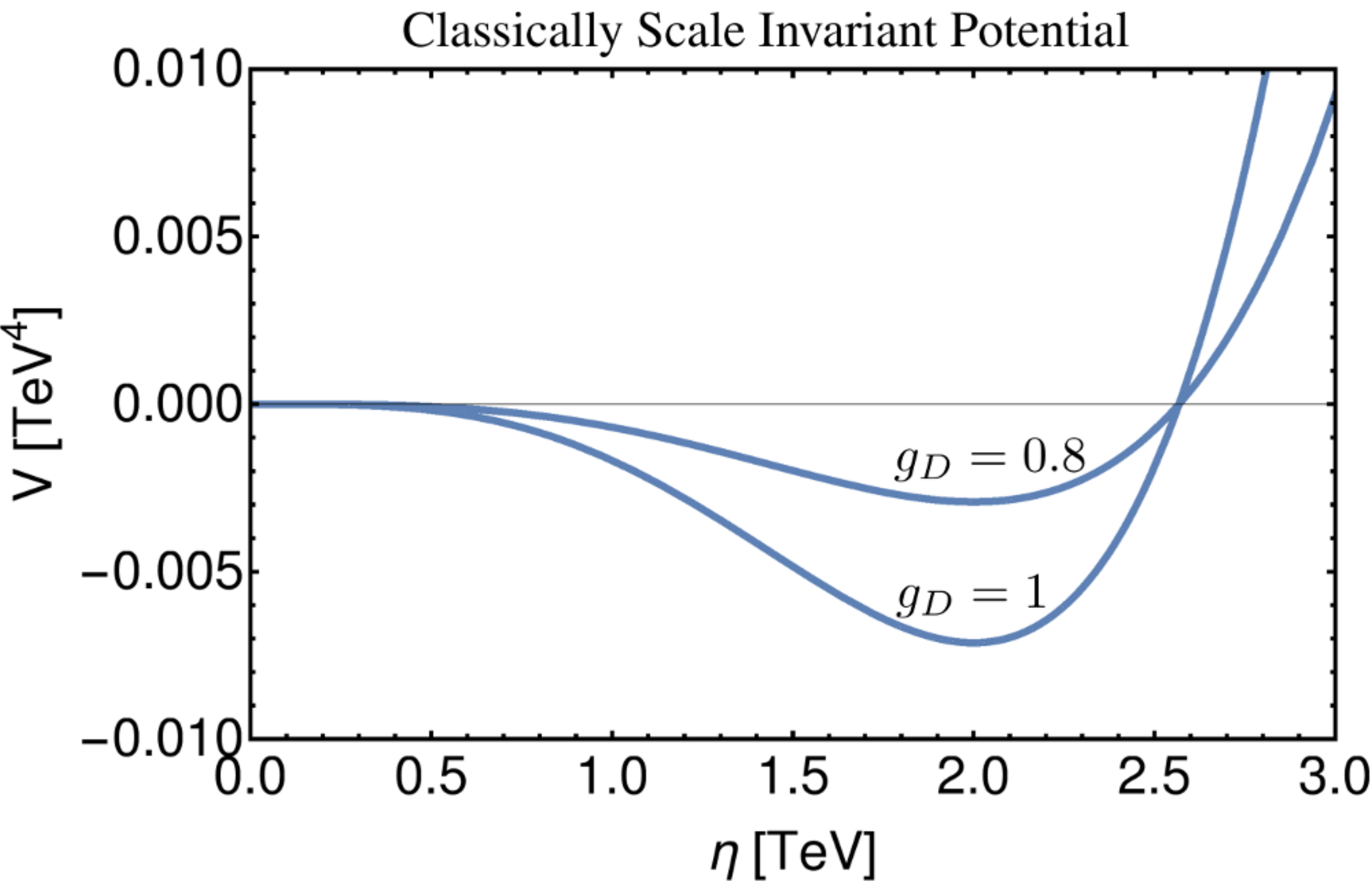}
\includegraphics[width=200pt,height=130pt]{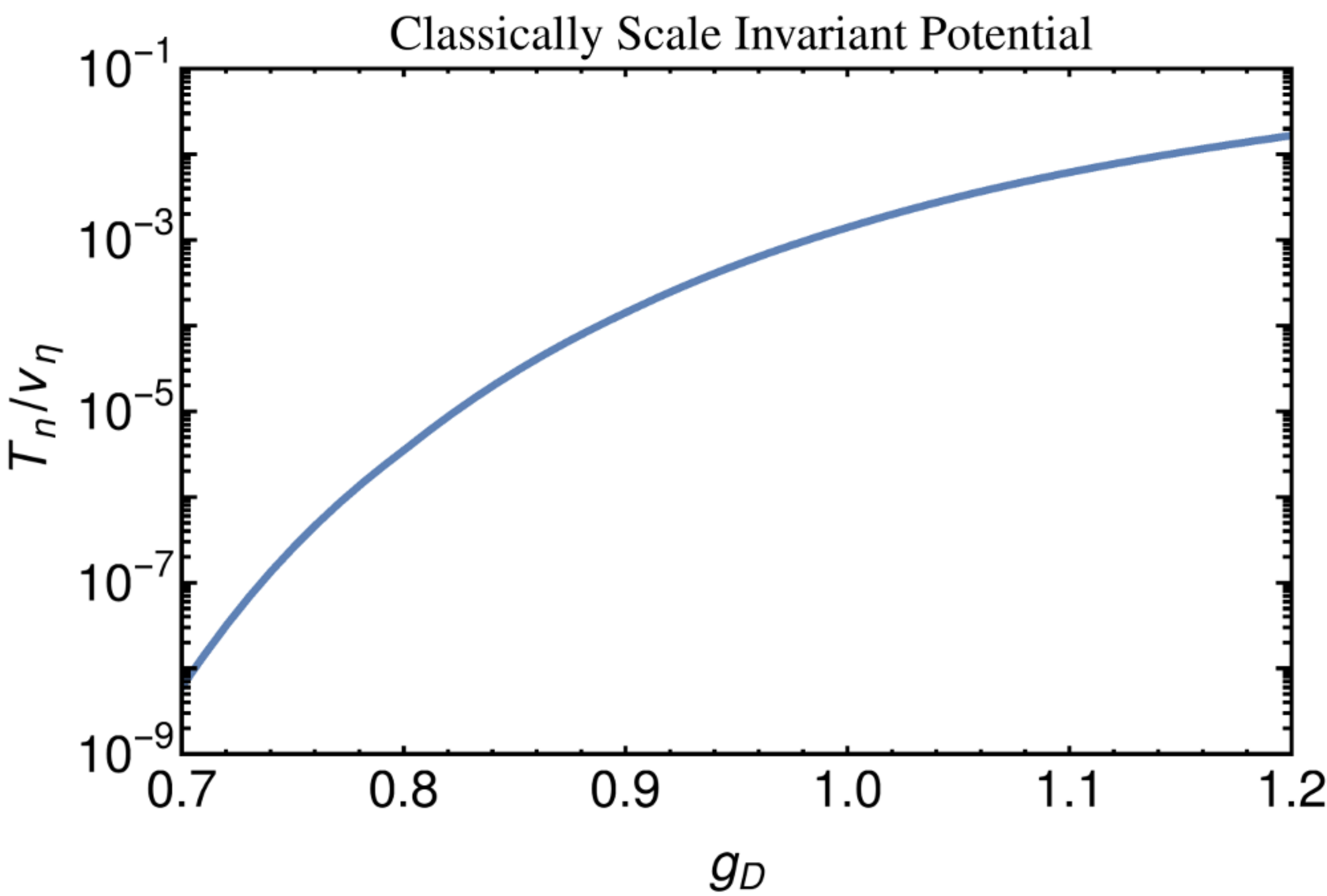}
\end{center}
\caption{\emph{Left:} the classically scale invariant potential ($T=0$) with $v_{\eta}=2$ TeV and two choices of the gauge coupling. \emph{Right:} the nucleation temperature as a function of the gauge coupling in the classically scale invariant case, for a fixed nucleation condition $S_{3}/T=142$, and ignoring QCD effects.}
\label{fig:scaleinv_potential}
\end{figure}

Another possibility is to impose classical scale invariance on the theory, as explained above.
This scenario with our field content has been studied in~\cite{Hambye:2013sna,Hambye:2018qjv}. Such a potential typically exhibits a large amount of supercooling~\cite{Witten:1980ez,Guth:1980zk,Buchmuller:1990ds,Kuzmin:1992up,Espinosa:2008kw,Konstandin:2011dr,Spolyar:2011nc,Dorsch:2014qpa,Jaeckel:2016jlh,Jinno:2016knw,Kubo:2016kpb,Kobakhidze:2017mru,Tsumura:2017knk,Marzola:2017jzl,Iso:2017uuu,Aoki:2017aws,Arunasalam:2017ajm,vonHarling:2017yew,Huang:2018fum}. This is because, lacking a mass term, the $T=0$ potential is very flat in field space. Furthermore, the positive thermal corrections from the gauge bosons will lead to a barrier being present for any finite $T$. Somewhat counter intuitively, a smaller $g_{D}$ actually leads to more supercooling because the $T=0$ potential becomes shallower, as shown in Fig.~\ref{fig:scaleinv_potential}. The thermal barrier also becomes smaller, but the shallower potential ends up being the more important effect. 

Of importance for the DM relic density in this scenario, is not just the DM annihilation cross section, but also the details of the phase transition. In particular the nucleation temperature, $T_{n}$, the temperature when inflation starts, $T_{\rm infl}$, and the reheating temperature, $T_{\rm RH}$. The latter two quantities are calculated following the methods in~\cite{Hambye:2018qjv}.

Furthermore, due to the large amount of supercooling, the PT may actually not take place before the temperature falls to $T \sim \Lambda_{\rm QCD}$. In this particular case, the $SU(2)_{D}$ PT is induced by QCD effects~\cite{Witten:1980ez,Buchmuller:1990ds,Iso:2017uuu,vonHarling:2017yew}. Our calculation of the nucleation temperature, ignoring the QCD trigger for now, is shown in Fig.~\ref{fig:scaleinv_potential}.

As a result, in the classical invariance scenario two distinct possibilities for the relic density can play out.
	\begin{itemize}
	\item \textbf{Regime (i): standard freeze-out.}
	\subitem{\underline{(ia). $T_{n} > \Lambda_{\rm QCD}$.}} 
	 There is a large thermal abundance of massive gauge bosons after the phase transition, i.e.~if $T_{\rm RH}/m_{A}$ and $g_{D}$ are large enough to bring (or keep) the gauge bosons in thermal equilibrium.
	Therefore, following the phase transition, the relic density is set through the usual freeze-out mechanism. Typically this occurs for gauge couplings $g_{D} \sim 1$ and $m_{A} \gtrsim 1.2$ TeV. 
	\subitem{\underline{(ib). $T_{n} < \Lambda_{\rm QCD}$.}} 
	This is similar to above, except the sequence of PTs is switched. Most of the parameter space corresponding to this regime has been ruled out by direct detection~\cite{Hambye:2018qjv}, except for the mass range $0.9 \; \mathrm{TeV} \lesssim m_{A} \lesssim 1.2$ TeV, see Fig.~\ref{fig:DD}.
	\item \textbf{Regime (ii): super-cool DM. }
	\subitem{\underline{(iia). $T_{n} > \Lambda_{\rm QCD}$.}} There is sufficient supercooling for a period of late time inflation to take place. Before the phase transition, the gauge bosons are massless and 	have a large abundance. This abundance is diluted away by the period of late time inflation.  The relic density in principle consists of the diluted, now super-cool, population of gauge bosons, together with an additional sub-thermal component created through scatterings after reheating. Numerically, however, we find the sub-thermal population is negligible in the parameter space corresponding to this regime, leaving the DM relic abundance set by the super-cool population of gauge bosons. The parameter space here corresponds to  $g_{D} \sim 1$ and $m_{A} \gtrsim 370$ TeV.
	\subitem{\underline{(iib). $T_{n} < \Lambda_{\rm QCD}$.}}  This is again conceptually similar to above except the PTs are switched. The sub-thermal DM population is now important for a large range of the parameter space, which corresponds to $g_{D} \lesssim 1$ and $m_{A} \lesssim 370$ TeV.
	\end{itemize}
In all regimes, once the relic density constraint is used, we are left with one free parameter which we take to be $m_{A}$. Here we wish to point out, supported by our calculations, that large portions of the parameter space of the classically scale invariant scenario can be probed through GW observatories. We shall now  in turn discuss the GW signal in regimes (ia) and (iia), which both exhibit promising GW signals. Regimes (ib) and (iib) on the other hand, which are less promising and include larger uncertainties, are relegated to appendix~\ref{sec:regb}.

\begin{figure}[t]
\begin{center}
\includegraphics[width=200pt]{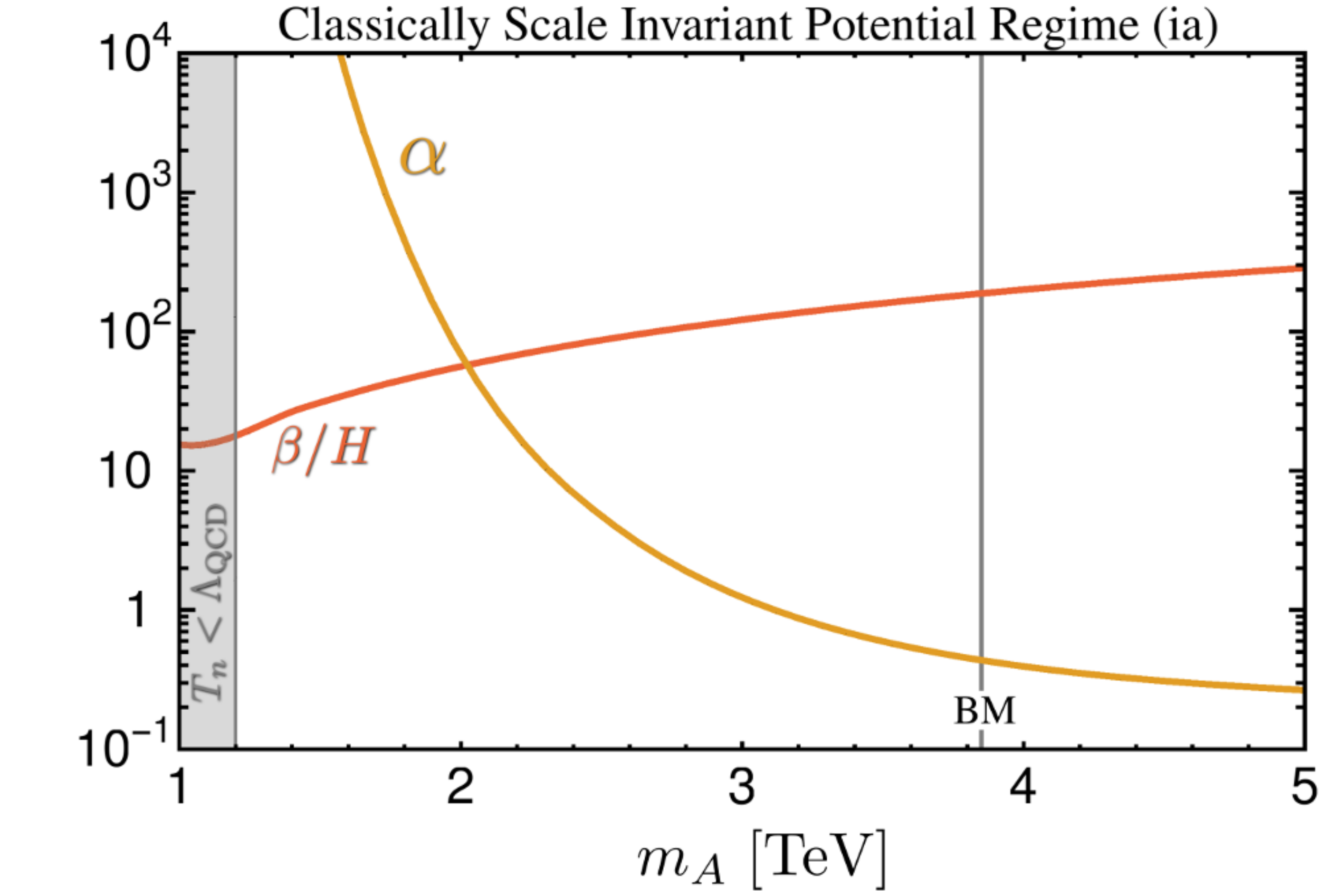}
\includegraphics[width=200pt,height=135pt]{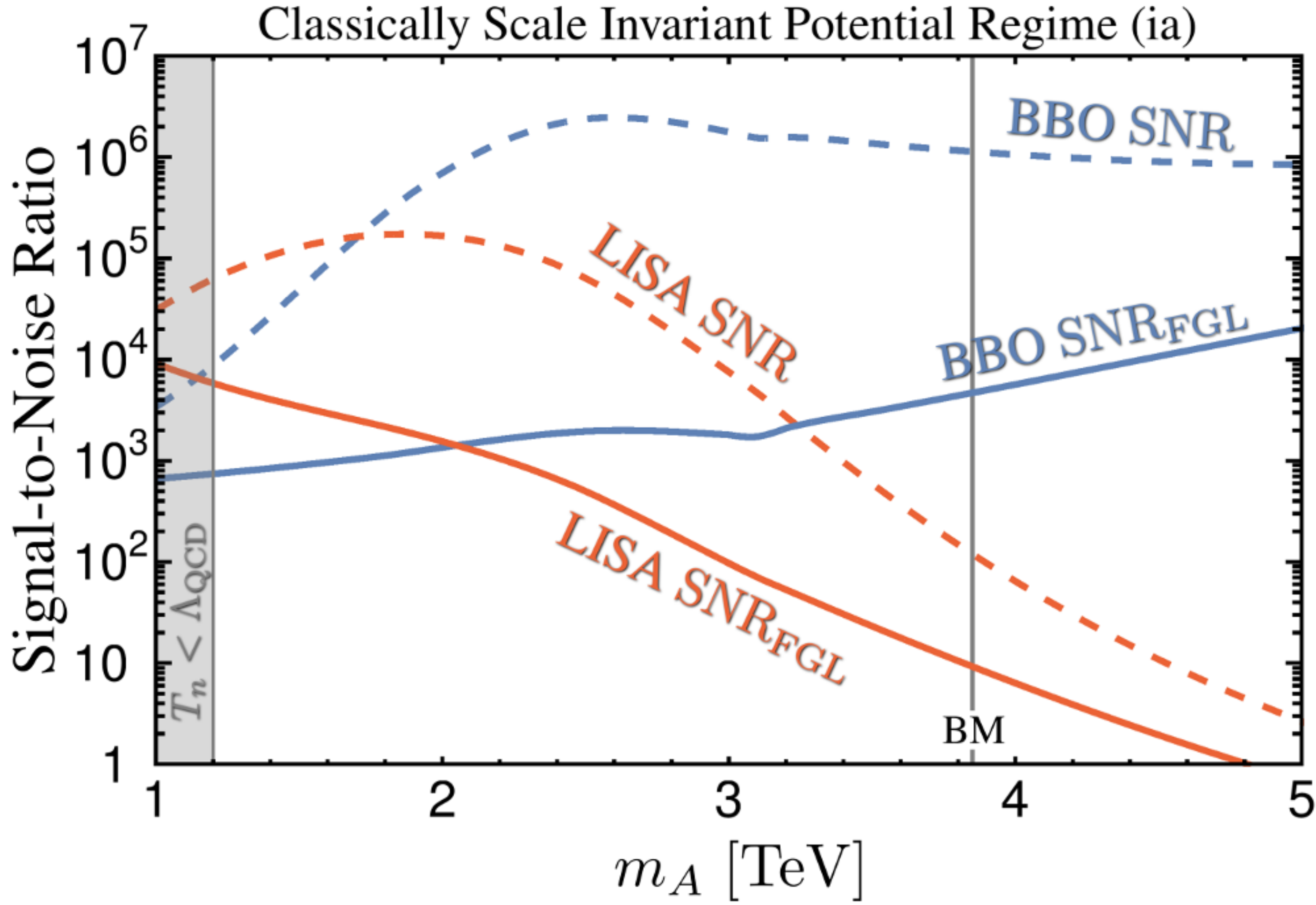}
\end{center}
\caption{\emph{Left:} the key phase transition parameters in regime~(ia) of the scale invariant case. \emph{Right:} The SNR for LISA and BBO. The Bodeker-Moore criterion, showing $v_{w} \simeq 1$, is satisfied for $m_{A} \lesssim 3.8$ TeV. Above this we still assume $v_{w} \approx 1$, though it could be lower, which would reduce the SNR.}
\label{fig:scaleinv_SNR}
\end{figure}

\begin{figure}[t]
\begin{center}
\includegraphics[width=200pt]{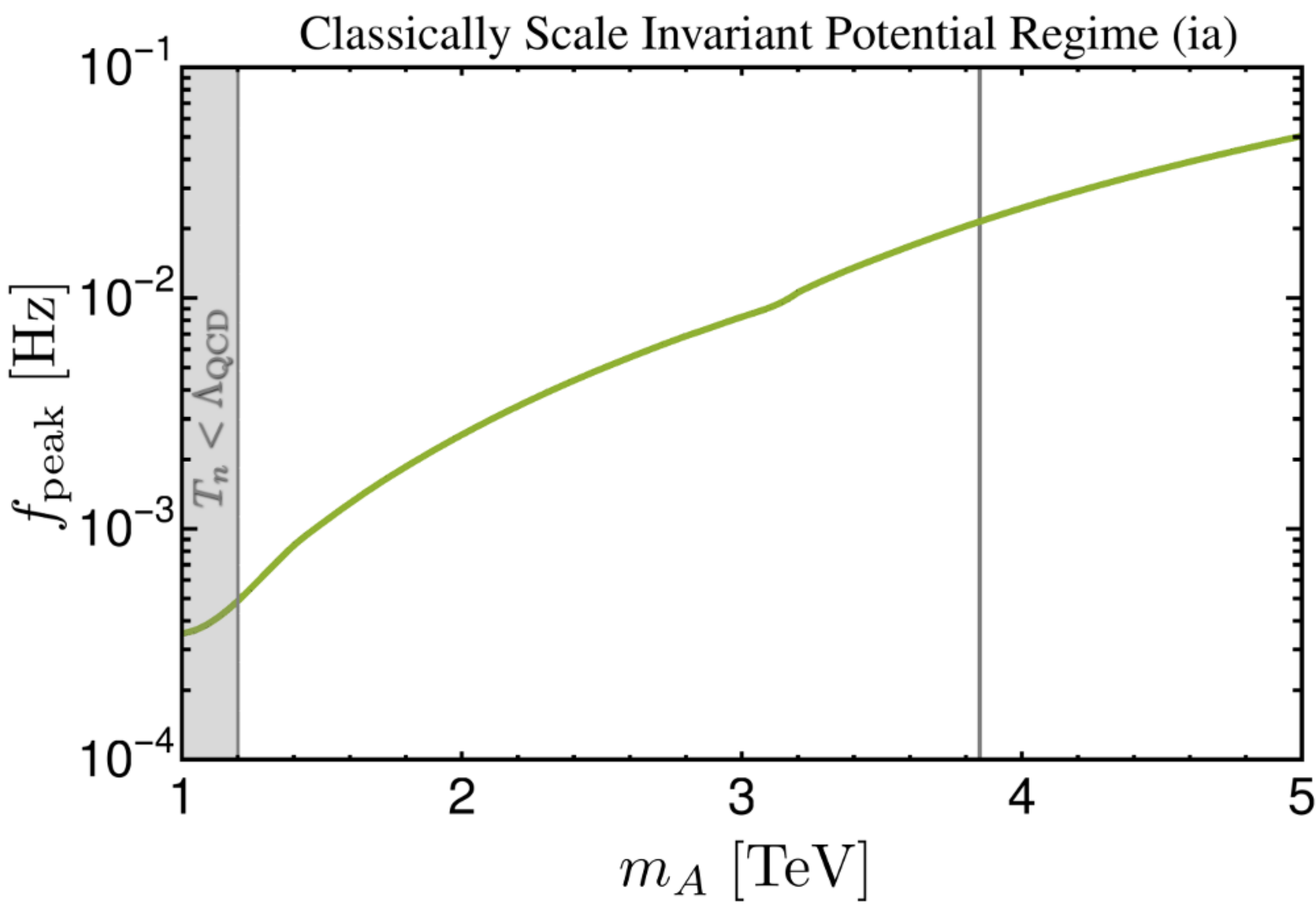}
\includegraphics[width=200pt,height=135pt]{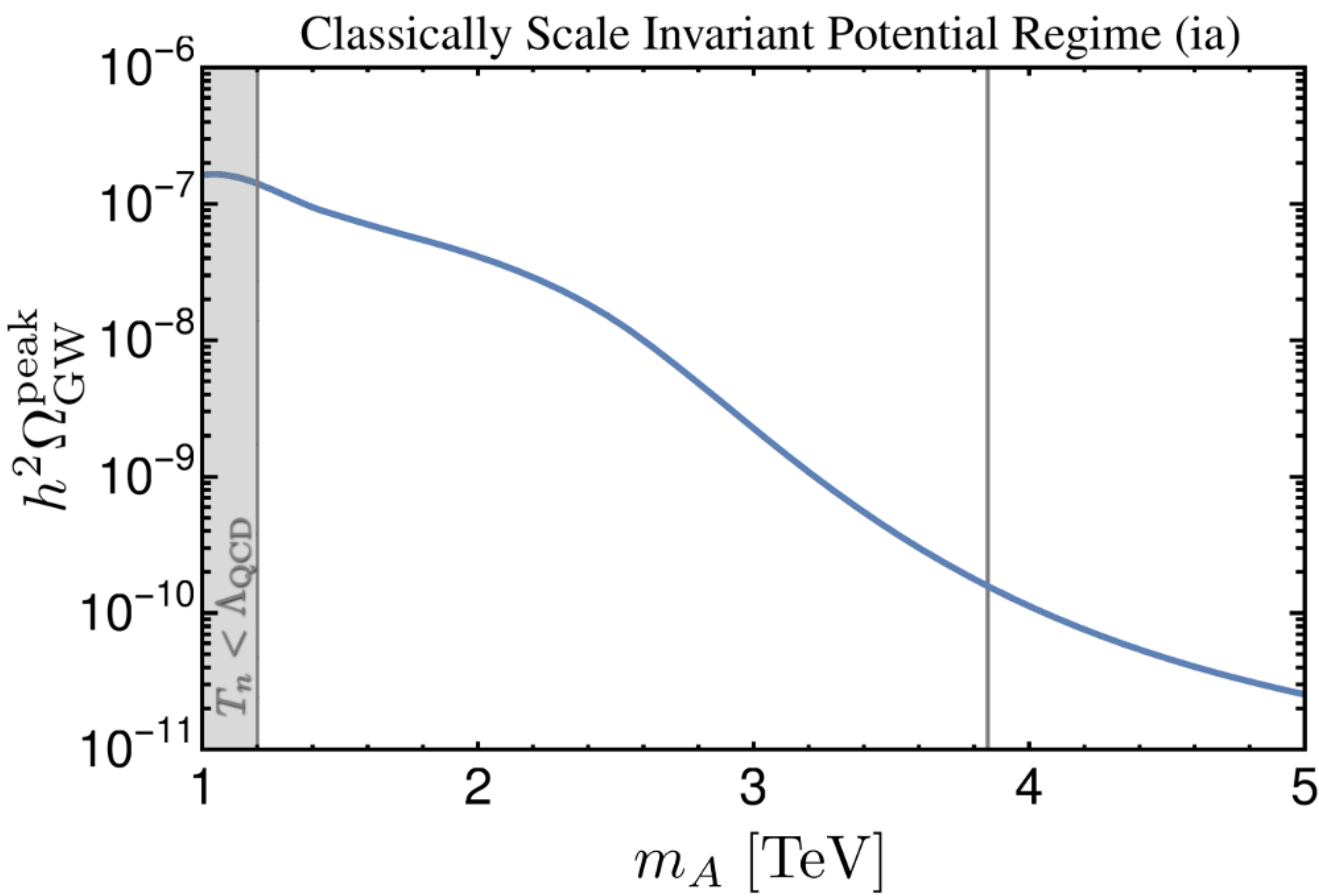}
\end{center}
\caption{\emph{Left:} the peak frequency in regime~(ia). \emph{Right:} the gravitational wave amplitude at the peak frequency in regime~(ia).}
\label{fig:reg1amore}
\end{figure}

\subsubsection*{Regime~(ia)}
The GW signal in this regime has previously been discussed in~\cite{Hambye:2013sna}. Here we provide our own --- updated and expanded --- calculation for completeness. 
For simplicity we assume the spectrum is given by the sum of the sound wave and turbulent contributions over the entire regime, although $H$ becomes vacuum dominated in the lower $m_{A}$ range. For justification of this choice, together with details of the GW spectrum used, see Appendix~\ref{sec:appB}. The key phase transition parameters are shown in Figs.~\ref{fig:scaleinv_SNR} and \ref{fig:reg1amore}, together with the foreground-free and foreground-limited signal-to-noise ratios. Note reheating is efficient in this regime: there is no period of matter domination immediately following the PT, as the decay rate of the inflaton is sufficiently large, $\Gamma > H$. 

As can be seen from Fig.~\ref{fig:scaleinv_SNR}, LISA can probe DM masses in this regime up to $m_{A} \sim 4$ TeV, even in the presence of the WD-WD foreground. This is more than competitive with projections for future direct detection experiments~\cite{Aprile:2015uzo,Akerib:2015cja,Mount:2017qzi,Aalbers:2016jon}, which can probe up to $m_{A} \sim 2$ TeV~\cite{Hambye:2018qjv}. (The current direct detection constraint demands $m_{A} \gtrsim 0.9$ TeV~\cite{Hambye:2018qjv,Aprile:2017iyp,Cui:2017nnn,Akerib:2016vxi}.)
The BBO proposal could test the entire parameter space shown here, well into what corresponds to the neutrino floor for direct detection experiments. Note for $m_{A} \lesssim 1.2$ TeV we find ourselves in regime~(ib), which is discussed below.

\subsubsection*{Regime~(iia)}

\begin{figure}[t]
\begin{center}
\includegraphics[width=200pt]{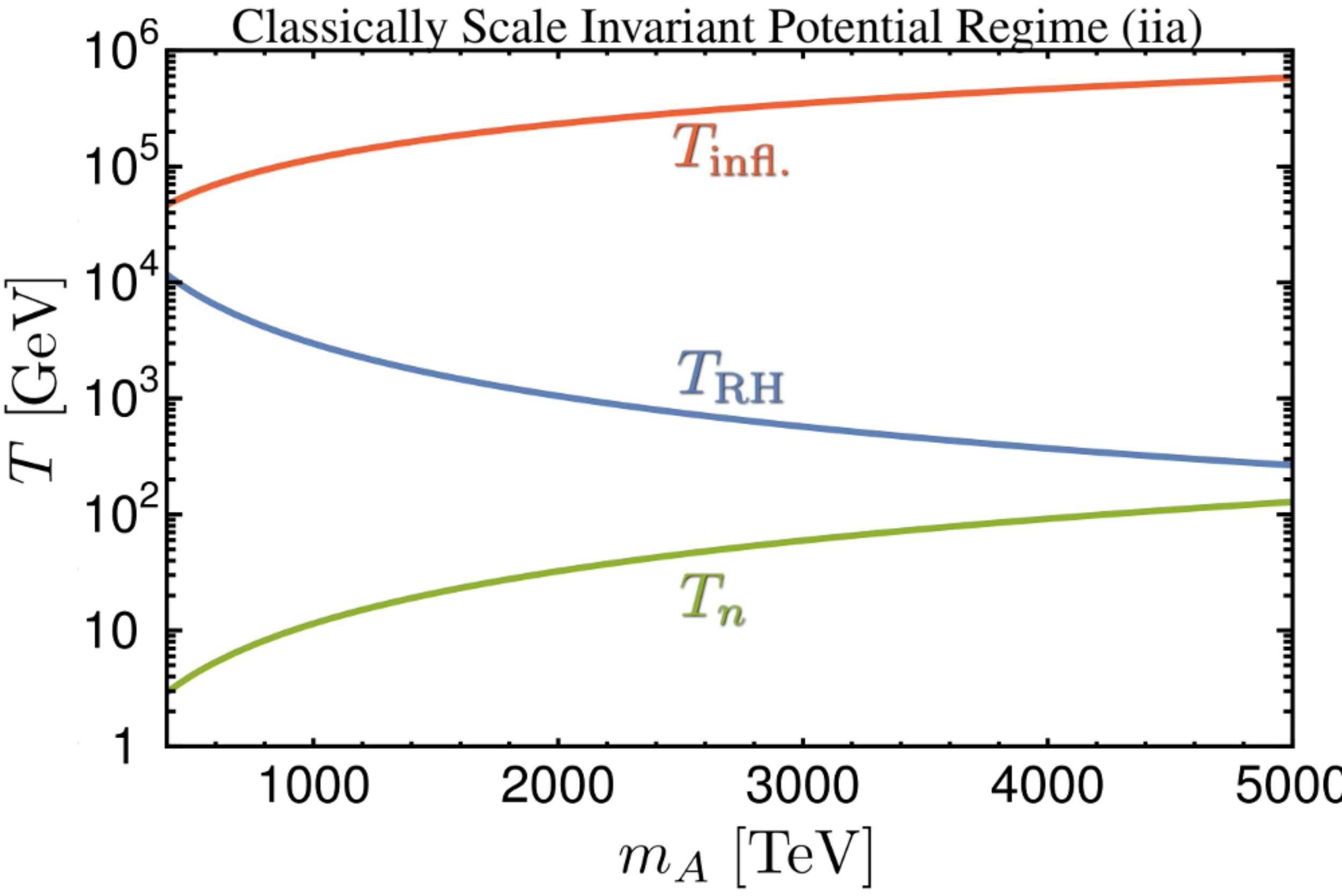}
\includegraphics[width=200pt]{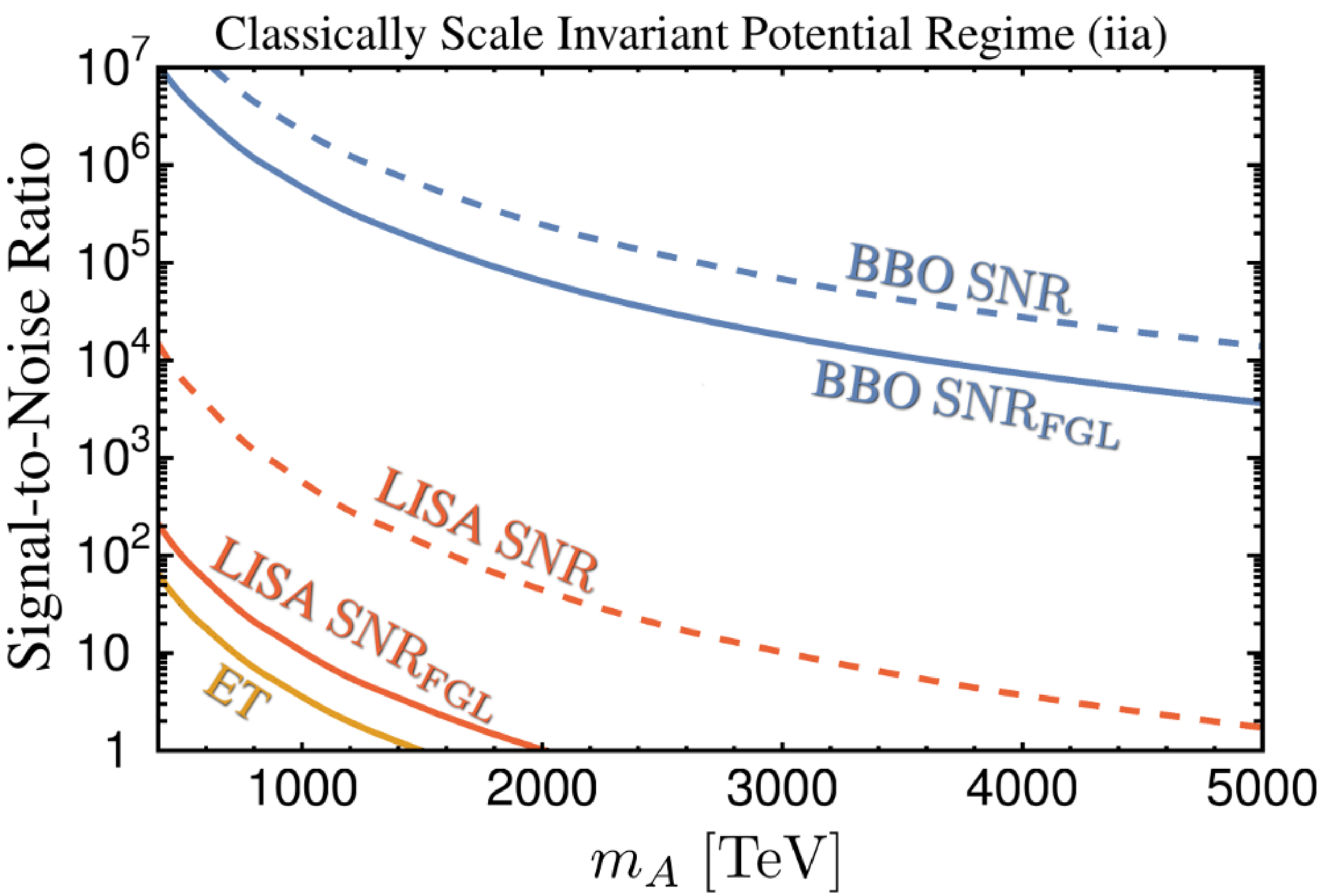}
\end{center}
\caption{\emph{Left:} The temperature when inflation starts, $T_{\rm infl}$, the reheating temperature, $T_{\rm RH}$, and the nucleation temperature, $T_{n}$, in regime (iia). The ratio $(T_{\rm infl}/T_{\rm RH})$ determines the amount of additional redshifting of the signal due to the matter dominated reheating period following the PT. \emph{Right:} the detectability of the GW signal in regime (iia). Here the BM criterion holds over the entire range.}
\label{fig:regimeiib_SNR}
\end{figure}

Following the methods in~\cite{Hambye:2018qjv}, we find this regime corresponds to parameters $g_{D} \approx 1$ and $m_{A} \gtrsim 370$ TeV. Notice that these DM masses are well above the usual unitarity constraint  from the thermal freeze-out of DM~\cite{Griest:1989wd,Baldes:2017gzw}, which does not take place  here. Numerically the required $g_{D}$ grows slowly, from $g_{D}=0.95$ for $m_{A} = 370$ TeV, to  $g_{D}=1.02$ for $m_{A} = 10000$ TeV. Our calculation of $T_{\rm RH}$ and $T_{\rm infl}$ is shown in Fig.~\ref{fig:regimeiib_SNR}. In this regime, reheating is inefficient following the PT, thus $T_{\rm RH}\neq T_{\rm infl}$. Indeed, there is a period of matter domination following the PT, as $\eta$ oscillates about the minimum of its potential. More precisely, the ratio of scale factors between the PT and the end of reheating is given by
	\begin{equation}
	\frac{ a_{\rm PT} }{ a_{\rm RH} } = \left( \frac{ T_{\rm RH} }{ T_{\rm infl} } \right)^{4/3}.
   \label{eq:ratiosa}
	\end{equation}
This leads to greater expansion of the Universe between the PT and today, suppressing the signal, and redshifting the frequency further than would otherwise be the case.

\begin{figure}[t]
\begin{center}
\includegraphics[width=200pt]{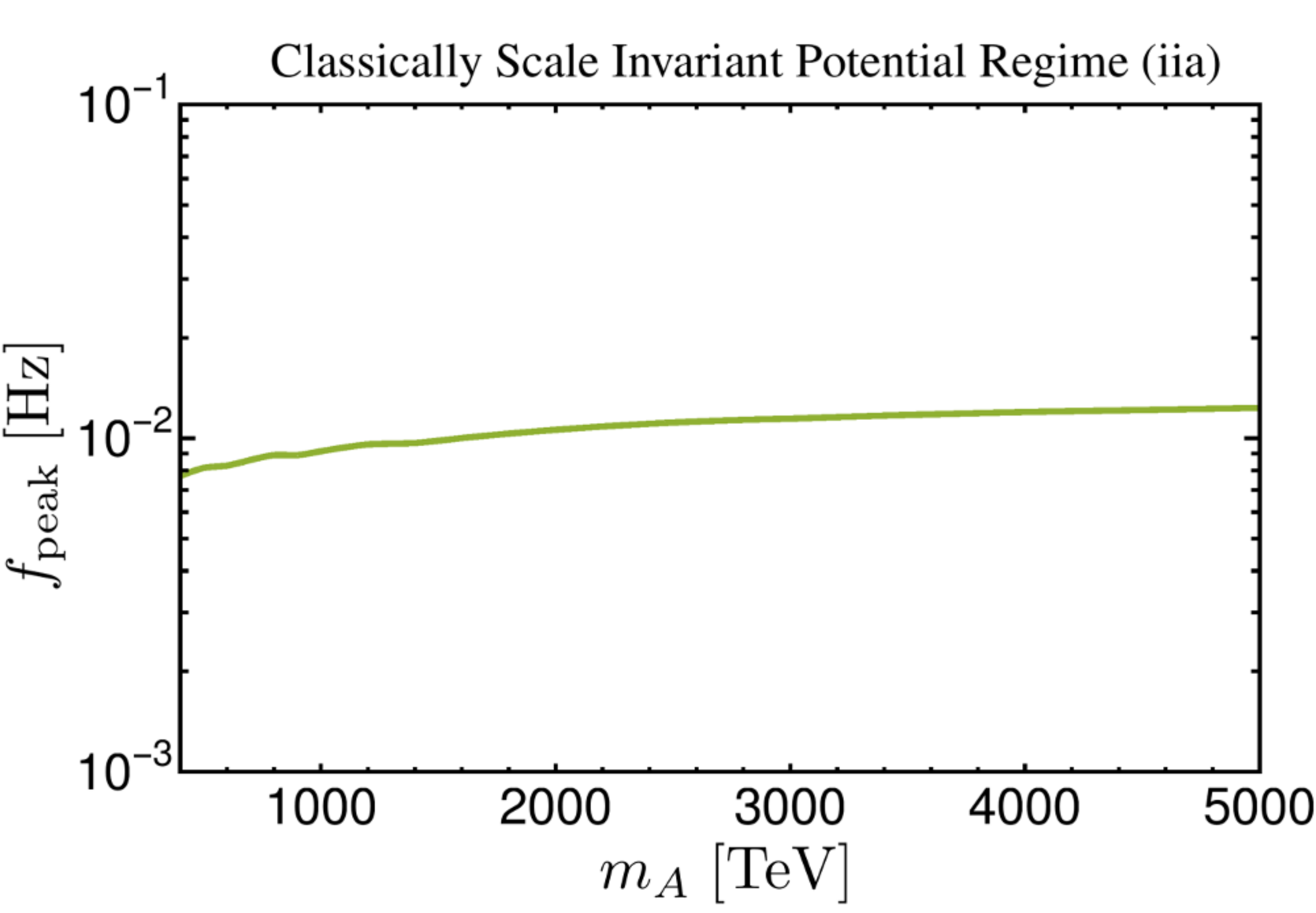}
\includegraphics[width=200pt]{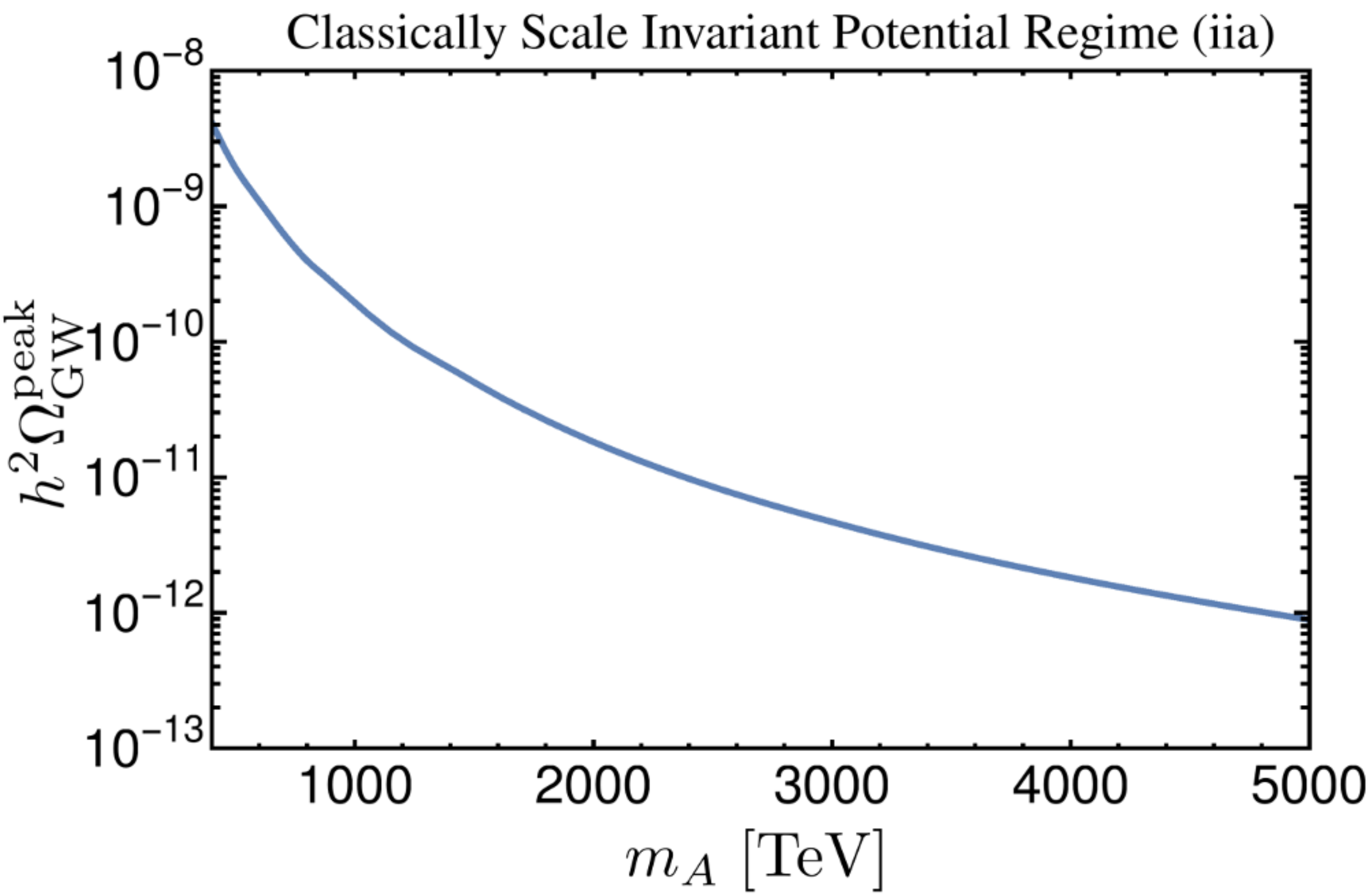}
\end{center}
\caption{\emph{Left:} The peak frequency in regime (iia). \emph{Right:} the peak amplitude in regime (iia). In this regime $\alpha \sim \mathcal{O}(10^{15})$ and $\beta /H \approx 7$ largely independent of $m_A$.}
\label{fig:regimeiia_more}
\end{figure}

\begin{figure}[t]
\centering
\includegraphics[width=0.55\textwidth]{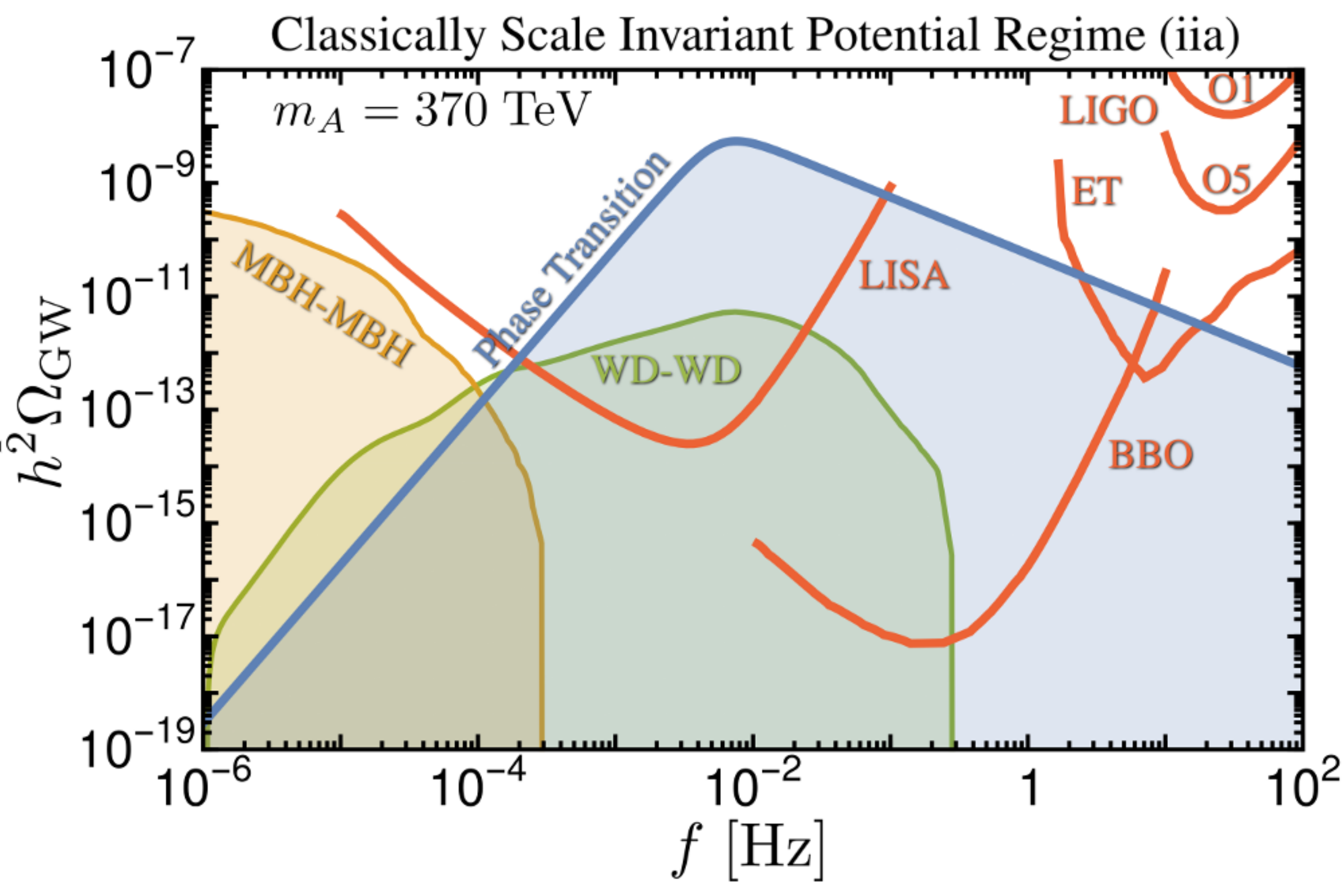}\hspace{30pt}
\raisebox{0.15\height}{\includegraphics[width=0.31\textwidth]{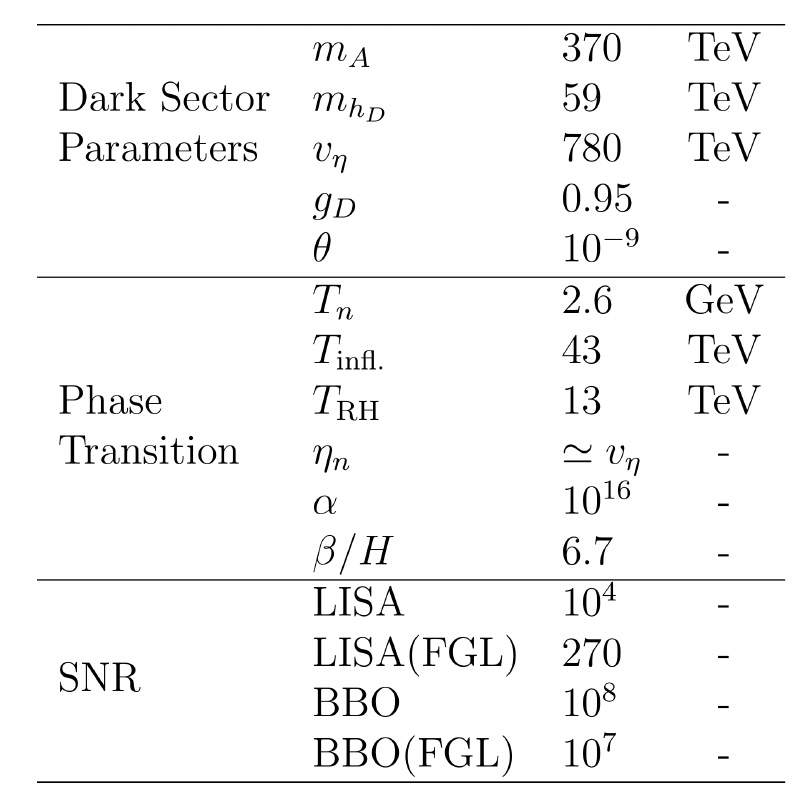}}\\
\includegraphics[width=0.55\textwidth]{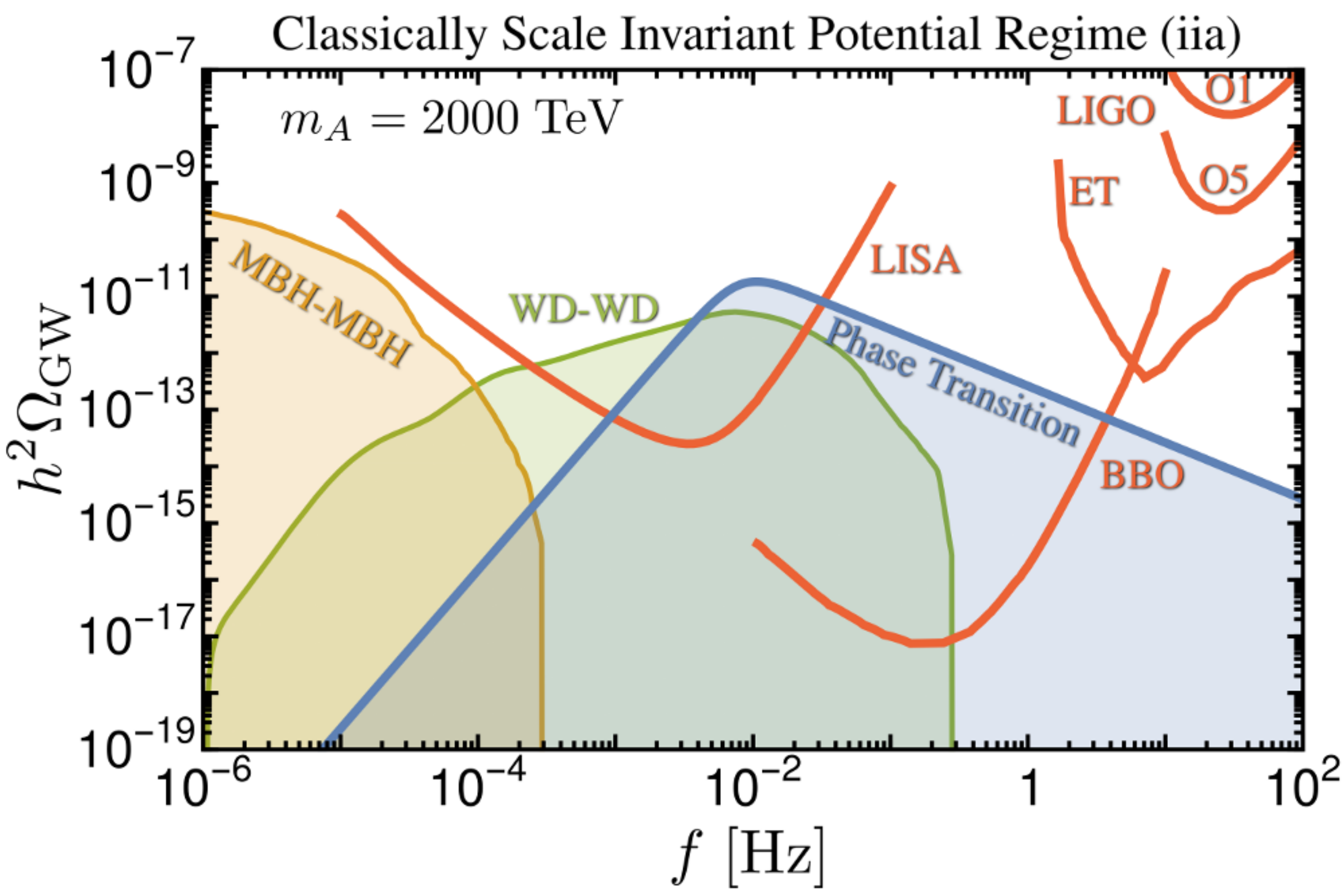}\hspace{30pt}
\raisebox{0.15\height}{\includegraphics[width=0.31\textwidth]{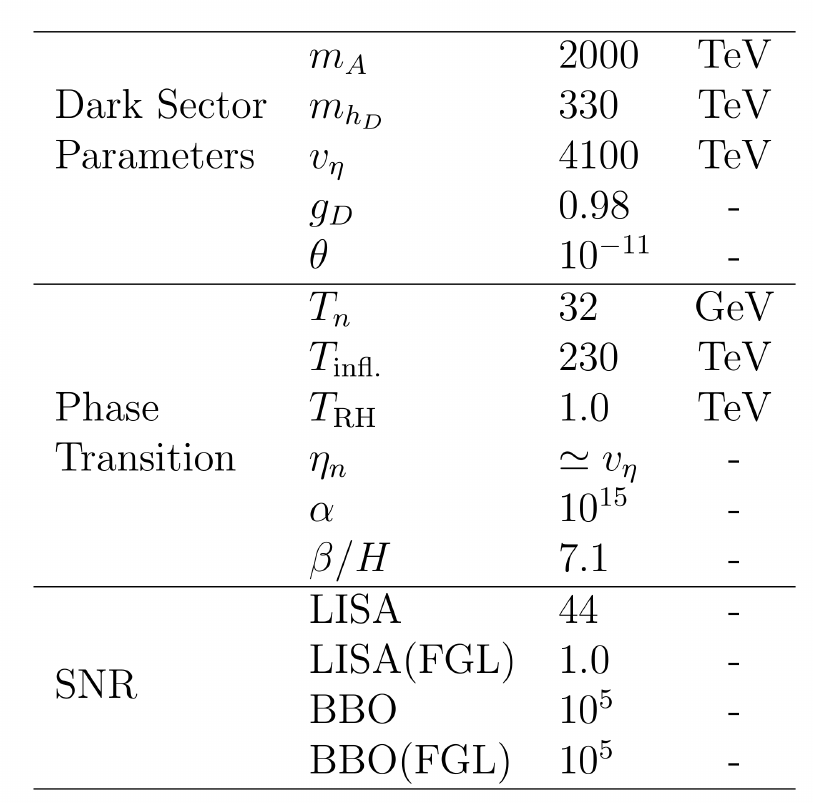}}\\
\caption{Examples of GW spectra in regime (iia). Although $\alpha \gg 1$, and $\beta/H$ is similar for both phase transitions, the period of matter domination after the PT is longer for larger $m_{A}$, leading to a suppressed signal. For purposes of illustration we also include the unresolvable foreground from black hole binaries with masses $10^{2}M_{\odot} - 10^{10}M_{\odot}$ (MBH-MBH)~\cite{Rosado:2011kv}.
}
\label{fig:regimeiib_spectra}
\end{figure}

The GW spectrum is determined in the following way. First of all, due to the large amount of supercooling the scalar field configuration --- and not sound waves or turbulence --- is the source of the signal (see Appendix~\ref{sec:appB2} for further discussion). It has been suggested that the oscillations of the scalar field after the PT may increase both that peak frequency and energy density of the GW signal by an order of magnitude~\cite{Child:2012qg}. We choose to remain conservative, however, and base our spectrum on the non-oscillating scalar field contribution, as indicated in Appendix~\ref{sec:appB1}. Once the Universe enters the late inflationary stage at $T_{\rm infl}$, the energy density remains constant until the plasma temperature reaches $T_{n}$, and so the Hubble scales at both temperature are the same $H(T_{n}) = H(T_{\rm infl})$. Taken together, $\beta$ and $H \sim T_{\rm infl}^{2}/M_{\rm Pl}$ set the initial frequency of the GW signal. We then redshift this value to $T_{\rm RH}$ when the Universe once again enters a radiation dominated phase. The redshifting from $T_{\rm RH}$ to today then follows the standard calculation~\cite{Caprini:2015zlo}. Taking all this, together with Eq.~\eqref{eq:ratiosa} into account, the peak frequency is given by
	\begin{equation}
	f_{\rm peak}^{(iia)} = 16.5 \, \mu\textrm{Hz} \left( \frac{ T_{\rm RH} }{ T_{\rm infl} } \right)^{1/3} \left( \frac{f_{\rm \ast} }{ \beta } \right) \left( \frac{ \beta} {H} \right) \left( \frac{ T_{\rm infl} }{ 100 \, \mathrm{GeV} } \right) \left( \frac{ g_{\ast} }{ 100 }\right)^{1/6},
\label{eq:re1}
	\end{equation}
where $g_{\ast}$ counts the effective degrees of freedom contributing to the radiation density, and $f_{\ast}/\beta =0.62/(1.8-0.1v_{w}+v_{w}^{2})$ is taken from simulations~\cite{Huber:2008hg}. Due to cancellations between the various factors, we find the peak frequency here is $\sim 10^{-2}$ Hz and almost independent of $m_A$, as shown in Fig.~\ref{fig:regimeiia_more}. The amplitude of the spectrum is also suppressed with respect to the case with no early period of matter domination,
	\begin{equation}
	\Omega_{\rm GW} \to \left( \frac{ T_{\rm RH} }{ T_{\rm infl} } \right)^{4/3}\Omega_{\rm GW},
\label{eq:re2}
	\end{equation}
because $\Omega_{\rm GW} = \rho_{\rm GW}/\rho_{c} \propto a^{0}, \, (a^{-1})$ in a radiation (matter) dominated Universe. Accounting for these factors, we find the GW spectra and summarise their detectability in Fig.~\ref{fig:regimeiib_SNR}. Examples of the spectra are shown in Fig.~\ref{fig:regimeiib_spectra}. 
Notice that for these large masses, the frequency of the gravitational waves extends well above 1 Hz, motivating us to compare our signal against sensitivity curves from current and future LIGO configurations O1 and O5~\cite{Abbott:2017xzg}, and ET~\cite{Punturo:2010zz,Hild:2010id,Regimbau:2011rp,Cui:2018rwi}. Finally note we have explicitly checked the phase transition completes even though we are in the vacuum dominated regime. Details are presented in Appendix~\ref{sec:appC}.

\section{Discussion and Conclusion}
\label{sec:conclusion}

We have explored the possibility of spin-one DM from a hidden $SU(2)_{D}$ gauge group. The stability of DM is elegantly assured through a custodial symmetry. Given the massive vector bosons, unitarity demands that the $SU(2)_{D}$ be broken through the Higgs mechanism. This implies a phase transition or crossover occurred in the dark sector, i.e.~the symmetry was initially unbroken at high temperatures. A strong phase transition will result in gravitational waves possibly detectable at future gravitational wave observatories. 

In this scenario the $SU(2)_{D}$ gauge coupling plays a crucial role in determining the relic abundance through freeze-out or late-time inflation. The same gauge coupling controls both the scattering cross-section and the thermal effects of the gauge bosons relevant for the phase transition. The model is therefore well suited as a case study for the sensitivity of future gravitational wave observatories to phase transitions in DM sectors.

We studied both tree level and radiatively-induced symmetry breaking. After finding the resulting gravitational wave spectra we identified parameter space which can be probed by LISA and BBO. As is known from previous studies, only limited parameter space of standard polynomial type potentials can be tested by LISA. The prospects improve for the classically scale invariant scenario. In this case, LISA is competitive with future direct detection experiments in the freeze-out regime and can probe the new regime of super-cool DM, which is inaccessible to direct and indirect detection.  Nevertheless, a conclusive test could only be performed by a more powerful observatory such as BBO.

We saw how foregrounds, which have so far been largely ignored in phase transition studies, apart from in~\cite{Hogan:1986qda,Grojean:2006bp}, can be taken into account in the estimates of the signal-to-noise ratio. Our results should be taken as indicative; we expect updated estimates of foregrounds to become available as our knowledge of the binary populations improves. More sophisticated studies, taking into account the precise capability of the LISA and eventually BBO spacecraft are required. Simulations of sound waves in the plasma for $\alpha > 0.1$ should also be performed. Only then will it be possible to conclusively rule out models from their implied gravitational wave signals using future LISA and BBO data. A positive signal at LISA --- which requires a very strong phase transition --- would most likely point toward exotic new physics at the TeV scale such as the close-to-conformal potential studied here.

\subsubsection*{Acknowledgements}
The authors thank D.~Teresi for helpful correspondence regarding super-cool DM. We thank J.~R.~Espinosa and T.~Konstandin for clarifying discussion. C.G.C. is supported by the ERC Starting Grant NewAve (638528).

\subsubsection*{Note Added} 
After our paper was released on arXiv, Ref.~\cite{Prokopec:2018tnq} appeared, which confirmed the validity of our one-dimensional field space approximation for the classically scale invariant potential.

\appendix

\section{The effective potential}
\label{sec:effectivepotential}

\subsection{Symmetry breaking at tree level}

The full effective potential is composed of four pieces

\begin{equation}
V(\phi,\eta,T)=V_{\rm tree} (\phi,\eta)+ V_{1}^{0}(\phi,\eta)+V^{\rm c.t}_{1}(\phi,\eta)+ V_{1}^{T}(\phi,\eta,T)\,.
\end{equation}

\subsubsection*{The tree-level piece}
This directly follows from Eq.~\eqref{Lag0} and it is given by
\begin{equation}
V_{\rm tree} (\phi,\eta) = \frac{\muh^2}{2} \phi^{2} + \frac{\lh}{4}\phi^{4} + \frac{\mueta^2}{2} \eta^{2}+ \frac{\leta}{4}\eta^{4} + \frac{\lheta}{4}\eta^{2} \phi^{2}.
\end{equation}

\subsubsection*{Coleman-Weinberg potential at zero temperature}
Knowing the field dependent masses, $m_{i}(\phi,\eta)$, in the Laudau gauge the one-loop $T=0$ contribution is given by
	
	\begin{align}
	V_{1}^{0}(\phi,\eta)= \sum_{i} \frac{g_{i}(-1)^{F}}{64\pi^2}  m_{i}^{4}(\phi,\eta)\left(\mathrm{Ln}\left[ \frac{m_{i}^{2}(\phi,\eta)}{\mu^2} \right] - C_i\right)\,,
	\label{eq:oneloop}
	\end{align}
where $\mu$ is the $\overline{\text{MS}}$ renormalization scale and   $g_{i}= \{1,3,6,12,1,9,3,3 \}$ for the $h, \; Z, \; W^{\pm}, \; t, \; \eta, \; A$, $G$, $G^D$. In addition, $C_i=5/2$ for  gauge bosons and $C_i=3/2$ otherwise. Finally, $F=0 \; (1)$ for bosons (fermions). 

The masses as a function of the scalar field values for the fermions and gauge bosons of the SM are 
\begin{align}
m_Z^2(\phi,\eta) = \frac{1}{4}(g_2^2+g_Y^2)\phi^2 \,,&&
m_W^2(\phi,\eta) = \frac{1}{4}g_Y^2\phi^2\,,&&
m_t^2(\phi,\eta) = \frac{1}{2} y_t^2 \phi^2\,.
\end{align}
Similarly, for the dark gauge bosons $m_A^2(\phi,\eta)= g_D^2 \eta^2/4$. Due to the coupling $\lambda_3$ in Eq.~\eqref{Lag0}, the scalar sectors mix with each other and the masses for the real scalar fields entering in Eq.~\eqref{eq:oneloop} are the eigenvalues of the matrix

\begin{eqnarray}
m^2_\text{Higgs}&=&
\left(
\begin{array}{cc}
\mu_1^2 + 3 \lh \phi^2 +\frac{1}{2} \lheta \eta^2 & \lheta\,  \phi \eta  \\
\lheta \,\phi \eta & \mu_2^2 + 3 \leta \eta^2 +\frac{1}{2} \lheta \phi^2 
\end{array}
\right)\,.
\label{eq:scalar_ms}
\end{eqnarray} 
In spite of this, the Goldstone bosons \emph{do not} mix at tree level. In fact, in the Landau gauge,  their masses are given by
\begin{eqnarray}
m_G^2(\phi,\eta) &=&  \mu_1^2 + \lambda_1 \phi^2+ \frac{1}{2}\lambda_3 \eta^2\,, \\
m_{G_D}^2(\phi,\eta) &=&  \mu_2^2 + \lambda_2 \eta^2+ \frac{1}{2}\lambda_3 \phi^2\,,
\end{eqnarray}
which vanish at $(\phi,\eta) = (v_\phi,v_\eta)$, as follows from Eqs.~\eqref{eq:mass_parameters}.

\subsubsection*{The counter-term potential} 
The counter terms to the potential in Eq.~\eqref{eq:oneloop} are 
\begin{equation}
V^{\rm c.t}_{1} (\phi,\eta) = \frac{\delta\muh^2}{2} \phi^{2} + \frac{\delta\lh}{4}\phi^{4} + \frac{\delta\mueta^2}{2} \eta^{2}+ \frac{\delta\leta}{4}\eta^{4} + \frac{\delta\lheta}{4}\eta^{2} \phi^{2}.
\label{eq:CT}
\end{equation}
By demanding no changes to the masses and VEVs of the scalars from their tree level values, that is, by imposing 
\begin{eqnarray}
\partial_{\phi,\eta} (V_{1}^{0} + V^{\rm c.t}_{1}) \bigg|_{(\phi,\eta)=(v_\phi,v_\eta)}&=& 0\,,\\
\partial_{\phi,\eta} \partial_{\phi,\eta} (V_{1}^{0} + V^{\rm c.t}_{1}) \bigg|_{(\phi,\eta)=(v_\phi,v_\eta)}&=& 0\,,\\
\end{eqnarray}
we calculate the couplings in Eq.~\eqref{eq:CT}. Moreover, we find\footnote{
There is a subtle issue for the contribution of the Goldstone bosons to Eq.~\eqref{eq:oneloop}. As explained above, their tree-level masses vanish at $(v_\phi,v_\eta)$ leading to an infrared divergence in Eq.~\eqref{eq:oneloop}. Such a divergence is spurious~\cite{Elias-Miro:2014pca} and disappears after accounting for the one-loop contributions to the Goldstone-boson self energies. Neglecting any possible mixing effect due to non-vanishing $\lambda_3$, the latter can be calculated by means of  
$
\delta m_G^2(\phi,\eta)=(1/2) \partial^2 V^{(0)}_1/\partial^2\phi $ and $
\delta m_{G_D}^2(\phi,\eta) = (1/2) \partial^2 V^{(0)}_1/\partial^2\eta$. 
}

\begin{eqnarray}
	V_{1}^{0}(\phi,\eta) + V^{\rm c.t}_{1}(\phi,\eta) &=& \sum_{i} \frac{g_{i}(-1)^{F}}{64\pi^2} \bigg\{  m_{i}^{4}(h,\eta)\left(\mathrm{Ln}\left[ \frac{m_{i}^{2}(\phi,\eta)}{m_{i}^{2}(v_\phi,v_\eta)} \right] - \frac{3}{2}\right) \nonumber \\
			    &&+2m_{i}^{2}(\phi,\eta)m_{i}^{2}(v_\phi,v_\eta)\bigg\} +{\cal O}(\lambda_3^2) \,.
	\end{eqnarray}
In this equation, the prescription for the scalars $m_i^2(\phi,\eta$) is the following. They are the eigenvalues of the mass matrix in Eq.~\eqref{eq:scalar_ms}, ordered in such way that
\begin{align}
m_\pm^2(0,0) = {\cal F}_\pm(\mu_1^2,\mu_2^2)\, && \text{and} &&
m_\pm^2(v_\phi,v_\eta) = {\cal F}_\pm(m_\phi^2,m_\eta^2)\,,
\end{align}
where
\begin{align}
{\cal F}_\pm(a,b)=\frac{1}{2}\left(a+b\pm(a-b)\text{sgn}(|a|-|b|)\right)\,.
\end{align}
Notice that $\Sigma_\pm {\cal F}_\pm(a,b)=a+b$ and that, when $a$ and $b$ are both positive (negative), ${\cal F}_+(a,b)$ is the maximum (minimum) of them.

\subsubsection*{Finite-temperature potential}
The one-loop finite $T$ contribution is given by
	\begin{align}
	V_{1}^{T}(\phi,T)= & \sum_{i} \frac{g_{i}(-1)^{F}T^{4}}{2\pi^2} \nonumber \\ & \times \int_{0}^{\infty}y^{2} \; \mathrm{Ln}\left(1-(-1)^{F}\mathrm{Exp}\left[-\sqrt{ y^{2}+{m_{i}^{2}(h,\eta) }/{ T^{2} } }\right]\right) dy.
	\label{eq:finiteT}
	\end{align}
We evaluate these integrals numerically.
In order to take into account the resummation of the Matsubara zero modes one includes the daisy term
	\begin{equation}
	V_{\rm Daisy}(\phi,T) = \sum_{i}\frac{\overline{g}_{i}T}{12 \pi}\Big \{ \big[m_{i}^{2}(\phi,\eta)\big]^{3/2}-\big[m_{i}^{2}(\phi,\eta)+\Pi_{i}(T)\big]^{3/2}\Big\}
	\label{eq:daisy}
	\end{equation}
where the sum runs only over scalars and the longitudinal degrees of freedom of the vector bosons, i.e $\bar{g_{i}} \equiv \{ 1,1,1,2,1,3,3,3 \}$ for $h, \; Z, \; \gamma, \; W^{\pm}, \; \eta, \; A,  \; G, \;G_D$. Here the thermal masses are given by~\cite{Carrington:1991hz}
	\begin{align}
	\Pi_\text{Higgs}		& = \begin{pmatrix}\frac{1}{2}\lh + \frac{1}{6}\lheta +\frac{3}{16}g_{2}^2 +\frac{1}{16}g_{Y}^2 + \frac{1}{4}y_{t}^{2}   & 0\\0 &\frac{1}{2}\leta + \frac{1}{6}\lheta +\frac{3}{16}g_{D}^2  \end{pmatrix} T^2\,, 
\\	
        \Pi_{G} 	& = \left(\frac{1}{2}\lh + \frac{1}{6}\lheta +\frac{3}{16}g_{2}^2 +\frac{1}{16}g_{Y}^2 + \frac{1}{4}y_{t}^{2}\right)T^2,\\
        \Pi_{G_D} 	& =\left( \frac{1}{2}\leta + \frac{1}{6}\lheta +\frac{3}{16}g_D^2 \right)T^2\,,\\
	\Pi_{Z/\gamma}		& =  \begin{pmatrix} \left(\dfrac{5}{6} +\dfrac{n_f}{3} \right)  g_{2}^2  & 0\\0 & \left(\dfrac{1}{6} +\dfrac{5n_f}{9} \right)g_{Y}^2  \end{pmatrix} T^2\,,\\
\Pi_{W}		& = \left(\dfrac{5}{6} +\dfrac{n_f}{3} \right)g_2^2T^{2}, \\
        \Pi_{A} 	& = \frac{5}{6}g_D^2T^{2},
	\end{align}
where $n_f=3$ is the number of fermionic families with $SU(2)\times U(1)$ charge.
Note for the scalars and the $Z/\gamma$, the prescription here is that $m_{i}^{2}(\phi,\eta)$ represents the relevant eigenvalue of the zero temperature mass matrix and $m_{i}^{2}(\phi,\eta)+\Pi_{i}(T)$ the relevant eigenvalue of the zero temperature mass matrix with the thermal masses added along the diagonal. This means the $Z$ and $\gamma$ mix at finite temperature. To avoid spurious contributions to the thermal masses from the $SU(2)_{D}$ gauge bosons at large field values, we cut off the $g_{D}$ contributions with a factor $(m_{A}/T)^{2}K_{2}(m_{A}/T)/2$, where $K_{2}(x)$ is the modified Bessel function of the
second kind of order two.

\subsection{Classically Scale Invariant Potential}
\label{sec:A2}
As explained in the text, in this case we have radiative symmetry breaking and the potential at one loop becomes~\cite{Coleman:1973jx,Gildener:1976ih,Hambye:2013sna,Hambye:2018qjv}
	\begin{equation}
	V_{1}^{0}(\eta) \simeq \frac{9g_{D}^{4} \eta^{4}}{512\pi^{2}}  \, \left(\mathrm{Ln}\left[ \frac{ \eta }{ v_{\eta} } \right] -\frac{1}{4}\right),
	\label{eq:CW}
	\end{equation}
where the $\phi$ direction plays a completely negligible role in the area of parameter space in which we shall be interested. (The EW symmetry is broken by the induced mass term, $\lambda_{3}v_{\eta}^{2}/2$, from the cross quartic.) The thermal effects are dominated by the gauge bosons. Thus the effective potential is well approximated by Eq.~(\ref{eq:CW}), together with the one-loop thermal, Eq.~(\ref{eq:finiteT}), and daisy terms, Eq.~(\ref{eq:daisy}), for the $SU(2)_{D}$ gauge bosons.

\section{The Gravitational Wave Spectrum}
\label{sec:appB}
\subsection{Summary of the Contributions}
\label{sec:appB1}
If directly after the PT the Universe becomes radiation-dominated, the stochastic GW background receives a number of contributions, summarised in~\cite{Caprini:2015zlo}. First, if no significant plasma is present, the scalar field contribution~\cite{Huber:2008hg} dominates 
\begin{eqnarray}
h^2\Omega_\text{GW}(f) & \simeq &    1.67 \times 10^{-5}   \, \left(\frac{H_*}{\beta} \right)^{2} \left( \frac{100}{g_*} \right)^{\frac{1}{3}} \left( \frac{\kappa_\eta \alpha}{1+\alpha} \right)^2  
\left(\frac{0.11\,v_w^3}{0.42+v_w^2}\right) \, S_1(f) .
\label{eq:scadom}
\end{eqnarray}
Alternatively, if a significant plasma is present, the following contributions dominate
\begin{eqnarray}
h^2\Omega_\text{GW}(f) &\simeq& 
 \underbrace{2.65 \times 10^{-6} \, \left(\frac{H_*}{\beta} \right)  \left( \frac{100}{g_*} \right)^{\frac{1}{3}} \left( \frac{\kappa_v \alpha}{1+\alpha} \right)^2  
  v_w \, S_2(f)}_{\text{sound waves}} \nonumber \\
& & +\underbrace{3.35 \times 10^{-4} \, \left(\frac{H_*}{\beta} \right)  \left( \frac{100}{g_*} \right)^{\frac{1}{3}}
\left(\frac{\kappa_{\rm turb}\,\alpha}{1+\alpha}\right)^{\frac{3}{2}}\,
 v_w \, S_3 (f)}_{\text{magnetohydrodynamic turbulance}} \,. \label{eq:plasmapresent}
\end{eqnarray}
Note we do not sum Eq.~\eqref{eq:scadom} with~\eqref{eq:plasmapresent}, following the updated recommendations in~\cite{Bodeker:2017cim}. Here $H_*$ is the Hubble parameter when the GWs are emitted and $\kappa_\eta$, $\kappa_v$, and $\kappa_{\rm turb}$ are the fractions of the latent heat that is converted into energy density in the scalar field, bulk motion of the fluid, and turbulence respectively. These can be calculated in terms of the wall velocity and $\alpha$. For this, we use the expressions reported in Ref.~\cite{Caprini:2015zlo,Espinosa:2010hh}. In addition, the spectral shapes are given by
\be
S_a( f ) =\left\{\begin{array}{ll}
3.8 \,\,(f/f_1)^{2.8}/\left[1 + 2.8 \, (f/f_1)^{3.8}\right] &a=1\\
 (f/f_2)^{3}\,\left[7/\left(4 + 3\,(f/f_2)^{2}\right) \right]^{\frac{7}{2}}&a=2\\
(f/f_3)^3/\left[\left( 1 + (f/f_3) \right)^{\frac{11}{3}} 
\left(1 + 8 \pi f/h_* \right)\right]  & a=3 \end{array} \right.\,.
\ee
The corresponding peak frequencies are
\be
\label{eq:peakb4}
f_a=\left(\frac{\beta}{H_*} \right) \left(\frac{T_*}{100\,{\rm GeV}}\right) \left( 
\frac{g_*}{100} \right)^{\frac{1}{6}} \left(\frac{1}{v_w}\right)\times \left\{\begin{array}{ll}16.5 \, {\rm \mu Hz}\,\times (f_{*}/\beta) & a=1 \\
 19\,{\rm \mu Hz}\,&a=2\\
27\,{\rm \mu Hz}  \,  & a=3
\end{array}
\right.\,,
\ee
where $f_{\ast}/\beta =0.62/(1.8-0.1v_{w}+v_{w}^{2})$ is taken from simulations~\cite{Huber:2008hg}. If the after the PT there is a period of matter domination, the previous expressions must be rescaled as explained in the main text. See Eqs.~\eqref{eq:re1} and ~\eqref{eq:re2}.

In this work, we are interested in phase transitions with significant supercooling, which lead to signals possibily observable by LISA. We caution that for strong phase transitions, $\alpha > 0.1$, considerable uncertainty enters into the use of Eqs.~(\ref{eq:scadom})-(\ref{eq:peakb4}), 
despite the fact  that they become insensitive to  $\alpha$ when the latter takes values much greater than one. This is partly due to the uncertainties in determining the relevant contribution, which have only been started to be explored following the updated results of~\cite{Bodeker:2017cim} (see appendix~\ref{sec:appB2} below). Furthermore, numerical simulations of sound waves in the plasma have also only been performed up to $\alpha = 0.1$. Until further simulations have been performed, we are therefore resigned to extrapolating the results to large $\alpha$, as in~\cite{Caprini:2015zlo}.

\subsection{Determination of the Relevant Contribution}
\label{sec:appB2}

We provide an estimate to justify our use of the sound wave plus magnetohydrodynamic turbulence in regime (ia), and the scalar field contribution in regime (iia), of the classically scale invariant potential. This is crucial as in the vacuum dominated regime, depending on the amount of supercooling, the energy transferred to the plasma and hence sound waves and magnetohydrodynamics may become negligible. 
Bodeker and Moore~\cite{Bodeker:2017cim} find a next-to-leading-order contribution to the pressure difference due to transition radiation across the wall which, using the parameters relevant to our model, scales as
	\begin{equation}
	\label{eq:NLO}
	\mathcal{P}_{\rm NLO} \sim \gamma g_D^2 m_A T^{3}. 
	\end{equation}
This is to be compared with the pressure driving the bubble expansion
	\begin{equation}
	\label{eq:expand}
	\mathcal{P}_{\rm Expand} = \Delta V \sim m_{A}^{4},	
	\end{equation}
where we assume vacuum domination and suppress some numerical factors in line with the accuracy of the analysis.
Now let us estimate whether the bubble wall continues to accelerate from nucleation until collision. At nucleation the bubble size is
	\begin{equation}
	r_{n} \sim \frac{1}{T}.
	\end{equation}
The bubble size at collision is given by the timescale of the transition, $\beta^{-1}$, multiplied by the wall velocity, $v_{w} \simeq 1$, hence 
	\begin{equation}
	r_{\rm coll} \sim \frac{1}{\beta}  \sim \left(\frac{H}{\beta}\right) \frac{M_{\rm Pl}}{m_{A}^{2}}.
	\end{equation}
Therefore, if the bubble continues to accelerate during its expansion, i.e. $\mathcal{P}_{\rm Expand} > \mathcal{P}_{\rm NLO}$, it will reach a highly relativistic state at collision
	\begin{equation}
	\label{eq:coll}
	\gamma_{\rm coll} = \frac{r_{\rm coll}}{r_{n}} \sim \left(\frac{H}{\beta}\right)\frac{M_{\rm Pl}T}{m_{A}^{2}},
	\end{equation}
typically $\sim 10^9$ in regime (iia). Using Eqs.~\eqref{eq:NLO}, \eqref{eq:expand}, and \eqref{eq:coll}, we find the condition $\mathcal{P}_{\rm Expand} > \mathcal{P}_{\rm NLO}$ is maintained throughout the expansion of the bubble, provided that
	\begin{equation}
	T_{n} \lesssim \left( \frac{\beta}{H} \frac{m_{A}^5}{g_{D}^2 M_{\rm Pl} } \right)^{1/4} = 0.5 \; \mathrm{TeV} \; \left( \frac{1}{ g_{D}} \right)^{1/2} \left( \frac{\beta}{H} \right)^{1/4}  \left( \frac{m_{A}}{ 1000 \; \mathrm{TeV}} \right)^{5/4}.
	\end{equation}
This is satisfied in regime (iia), but not in regime (ia), of our analysis. Thus in regime (iia) we expect a significant amount of energy to be stored in the walls --- rather than in the plasma --- before the bubbles collide. We wish to emphasise that the estimates provided here, similar to those in~\cite{Ellis:2018mja}, are of a preliminary nature. Indeed the transition in regime (iia) ends with an oscillating scalar field dominating the Universe, rather than with a relativistic plasma. The scalar field oscillations may further enhance the GW spectrum, as indicated by the results of numerical simulations~\cite{Child:2012qg}.

\section{Super-cool DM Regimes~(ib) and (iib)}
\label{sec:regb}
In this appendix we comment briefly on these regimes, in which the $SU(2)_D$ phase transition occurs after QCD confinement, although they are less promising from the point of view of GWs. In these regimes the QCD phase transition occurs with six massless quarks and is first order~\cite{Pisarski:1983ms,Brown:1990ev}. There is a chance this could lead to an observable GW signal~\cite{Schwaller:2015tja,Iso:2017uuu} (although note the vacuum domination continues for some time after the QCD PT, diluting the signal). We cannot, however, use our techniques above to accurately calculate the phase transition parameters, $\alpha$ and $\beta$ for the QCD phase transition~\cite{Iso:2017uuu}. Most likely a lattice study is required in order to more carefully explore this possibility. Alternative techniques have been pursued in~\cite{Tsumura:2017knk,Aoki:2017aws}.

We now turn to the details of the $SU(2)_{D}$ phase transition following after the QCD one. The quark condensate formed after chiral symmetry breaking leads to a tadpole term and hence a VEV for the EW Higgs. This in turn leads to a mass term for $\eta$ through the cross-quartic. Provided $3m_{A}^{2} \gtrsim 2m_{h}^{2}$, which corresponds to our regime of interest, the thermal barrier from the gauge bosons is still large enough to prevent immediate $SU(2)_{D}$  breaking. Instead, a first order phase transition occurs just before the barrier disappears at $T \sim m_{h}\Lambda_{\rm QCD}/m_{A}$. As can be checked numerically, the non-zero mass term for $\eta$ means this phase transition now occurs with a very large $\beta$, and does not lead to an observable gravitational wave signal.

\section{Completion of the Phase Transition}
\label{sec:appC}

\begin{figure}[t]
\begin{center}
\includegraphics[width=200pt]{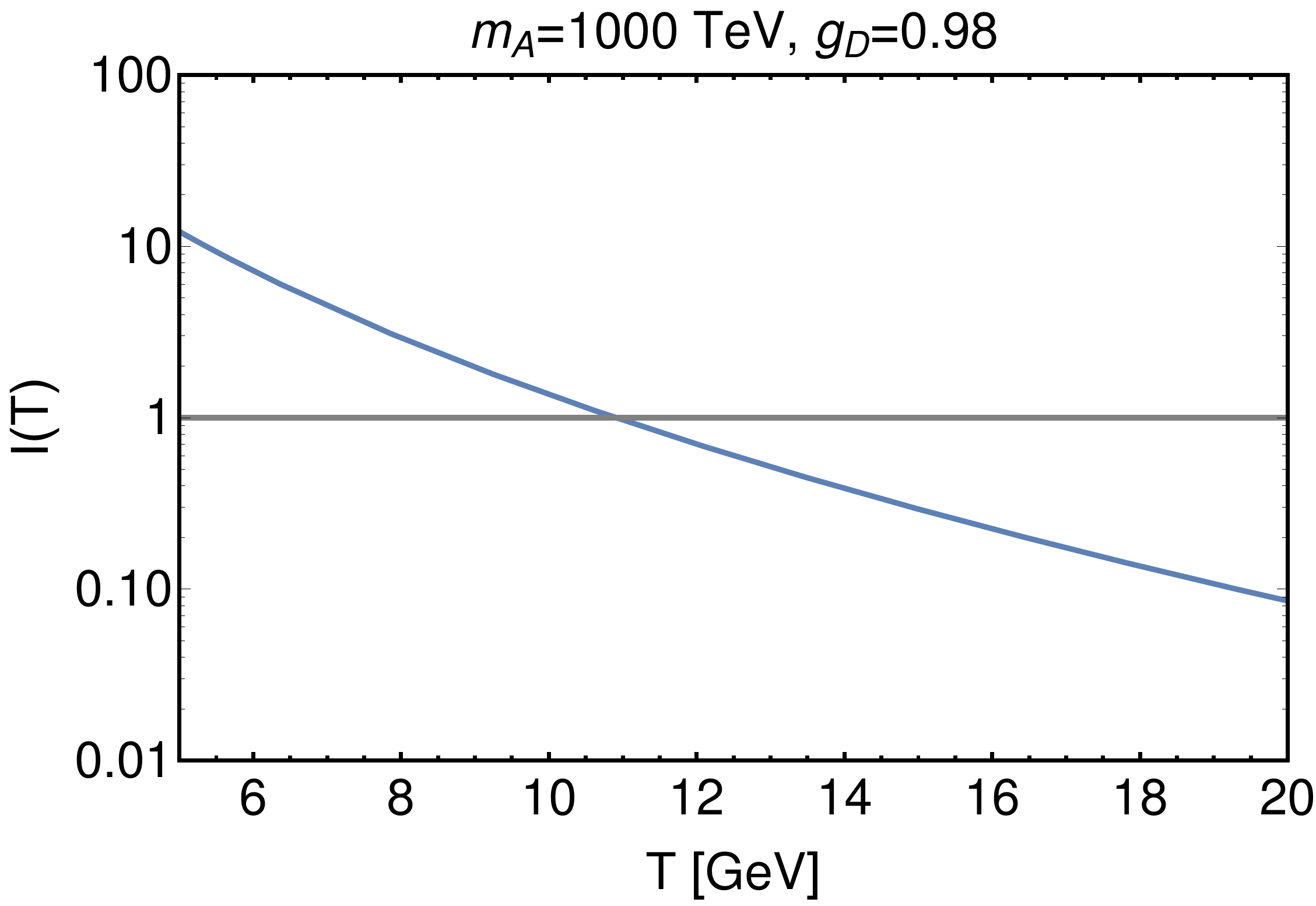}
\includegraphics[width=200pt,height=138pt]{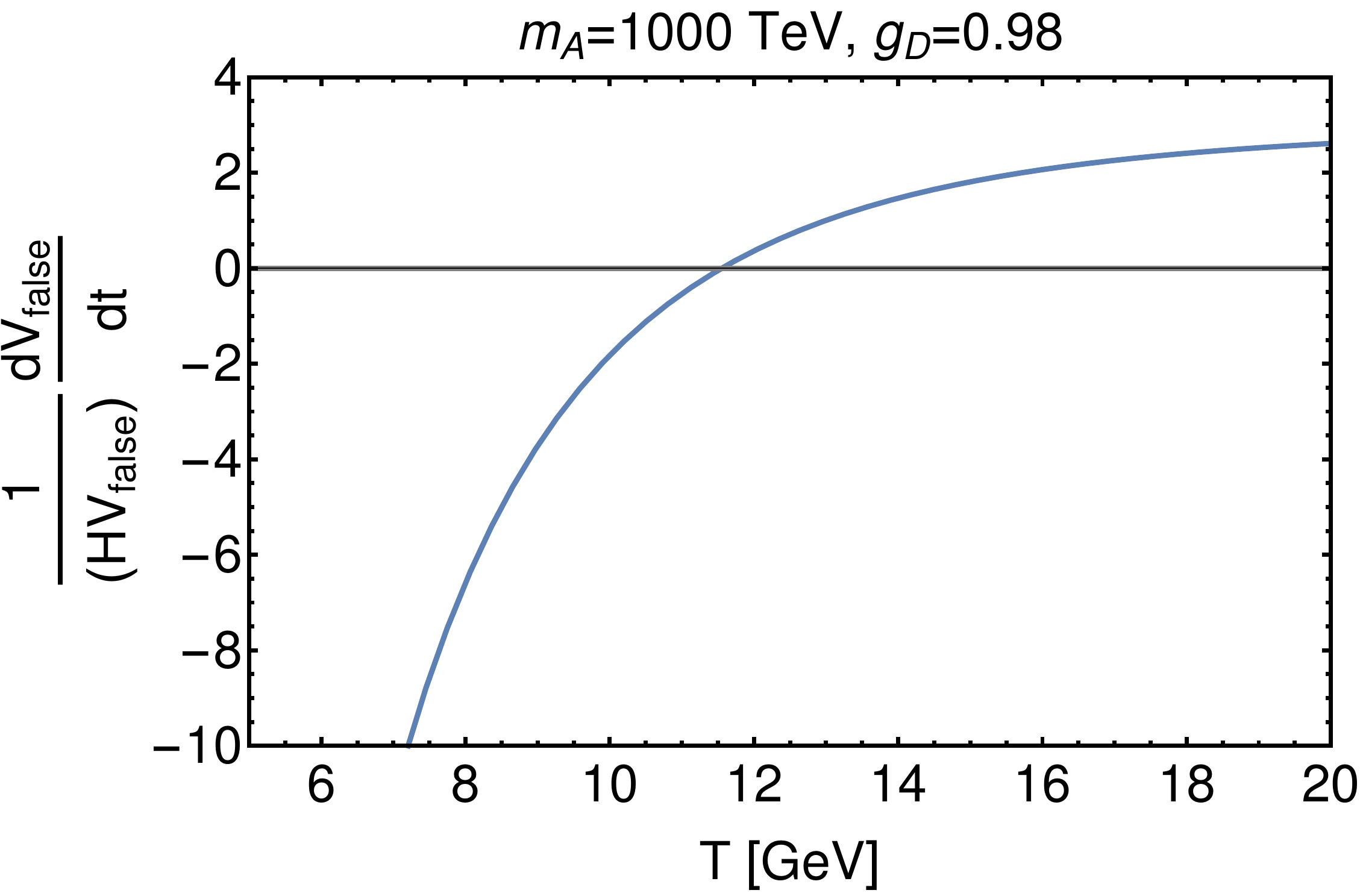}
\end{center}
\caption{Left: An example calculation for the phase transition in the classically scale invariant case. Here the percolation condition is achieved at $T_p \simeq 10.5$ GeV. Right: The physical volume of false vacuum is rapidly diminished at and below $T_p \simeq 10.5$ GeV.}
\label{fig:completion}
\end{figure}

We have checked the phase transitions in the classically scale invariant potential occurring in the vacuum dominated regime do indeed complete. This can be seen through an explicit calculation of the percolation temperature. Percolation requires a small probability of a point in the comoving volume being in the false vacuum~\cite{Guth:1979bh,Guth:1981uk,Guth:1982pn,Enqvist:1991xw,Ellis:2018mja}:
	\begin{equation}
	P(T) \equiv e^{-I(T)} \lesssim 1/e \implies I(T) \gtrsim 1,
	\end{equation}
where 
	\begin{equation}
	I(T) = \frac{4\pi}{3}\int_{T}^{T_c}dT'\frac{ \Gamma(T') }{ (T'H(T'))^{4} }\left(\int_{T}^{T'} \frac{ d\tilde{T} }{ H(\tilde{T}) } \right)^{3}.
	\end{equation}
One also requires that the physical volume of the false vacuum be decreasing significantly inside of one Hubble time~\cite{Guth:1979bh,Guth:1981uk,Guth:1982pn,Enqvist:1991xw,Ellis:2018mja}
	\begin{equation}
	\frac{1}{H \mathcal{V}_{\rm false}} \frac{d\mathcal{V}_{\rm false}} {dt} = 3 + T \frac{dI} {dT} \lesssim -1.
	\end{equation}
An example calculation showing these conditions are met is shown in Fig.~\ref{fig:completion}. The underlying reason this can occur is because the nucleation rate continues to grow exponentially as the temperature falls, allowing sufficient bubble formation to overcome the Hubble driven expansion of the false vacuum. In contrast, for very strong phase transtions in standard polynomial type potentials, $S_{3}/T$ may remain constant or even grow as the temperature drops. Meaning the phase transition may not complete even if the nucleation condition $\Gamma \sim H^{4}$ is achieved~\cite{Guth:1979bh,Guth:1981uk,Guth:1982pn,Enqvist:1991xw,Ellis:2018mja}.


\providecommand{\href}[2]{#2}\begingroup\raggedright\endgroup

\end{document}